\documentclass[10pt]{article}
\DeclareUnicodeCharacter{0301}{\'{e}}
\usepackage[utf8]{inputenc}
\usepackage[T1]{fontenc}
\usepackage{authblk}
\usepackage[svgnames,table]{xcolor} 
\usepackage{hyperref}
\usepackage{amsmath}
\usepackage{amssymb} 
\usepackage[utf8]{inputenc}
\usepackage{graphicx}
\usepackage{subcaption}
\usepackage{float} 
\usepackage{caption}
\usepackage[backend=biber,style=numeric,sorting=none]{biblatex} 
\usepackage{geometry}
\usepackage{tabularx,tabulary}
\title{Denoising Complex Covariance Matrices with Hybrid ResNet and Random Matrix Theory: Cryptocurrency Portfolio Applications}
\date{} % leave empty for no date

\author{Andrés García-Medina\thanks{Email: andgarm.n@gmail.com} \\ 
Faculty of Sciences, Autonomous University of Baja California, Ensenada, 22860, Mexico}

\begin{document}

\flushbottom
\maketitle

\thispagestyle{empty}

%% Abstract
\begin{abstract}
Covariance matrices estimated from short, noisy, and non-Gaussian financial time series are notoriously unstable. Empirical evidence suggests that such covariance structures often exhibit power-law scaling, reflecting complex, hierarchical interactions among assets. Motivated by this observation, we introduce a power-law covariance model to characterize collective market dynamics and propose a hybrid estimator that integrates Random Matrix Theory (RMT) with deep Residual Neural Networks (ResNets). The RMT component regularizes the eigenvalue spectrum in high-dimensional noisy settings, while the ResNet learns data-driven corrections that recover latent structural dependencies encoded in the eigenvectors. Monte Carlo simulations show that the proposed ResNet-based estimators consistently minimize both Frobenius and minimum-variance losses across a range of population covariance models. Empirical experiments on 89 cryptocurrencies over the period 2020–2025, using a training window ending at the local Bitcoin peak in November 2021 and testing through the subsequent bear market, demonstrate that a two-step estimator combining hierarchical filtering with ResNet corrections produces the most profitable and well-balanced portfolios, remaining robust across market regime shifts. Beyond finance, the proposed hybrid framework applies broadly to high-dimensional systems described by low-rank deformations of Wishart ensembles, where incorporating eigenvector information enables the detection of multiscale and hierarchical structure that is inaccessible to purely eigenvalue-based methods.
\end{abstract}



\section{Introduction}

Random Matrix Theory (RMT) has played a pivotal role in elucidating complex interactions in nuclear physics~\cite{mehta2004random}, revealing signatures of quantum chaos~\cite{bohigas1984characterization}, and describing quantum transport phenomena~\cite{beenakker1997random}. The Wishart distribution describes the sample covariance matrix of multivariate Gaussian data~\cite{wishart1928generalised}. Within the framework of RMT, such covariance matrices are commonly referred to as members of the Wishart Orthogonal Ensemble (WOE). The Wishart distribution can be viewed as a natural multivariate generalization of the relationship between the univariate normal distribution and the chi-square distribution~\cite{pena2002analisis}.

Wishart-type random matrix models naturally arise in a wide range of physical and applied contexts, including antenna selection in communication technologies~\cite{sadek2007active}, nuclear physics~\cite{fyodorov1997statistics}, quantum chromodynamics~\cite{verbaarschot1994spectrum}, and nonintersecting Brownian motions~\cite{schehr2008exact}.

More recently, RMT has been successfully applied to the denoising of empirical covariance matrices of Wishart type, leading to improved optimal asset allocation within portfolio theory~\cite{laloux1999noise, ledoit2011eigenvectors, bun2017cleaning, ledoit2020analytical}. Within this framework, several methodologies have been proposed to identify meaningful signals and suppress noise arising from the finite size of empirical samples~\cite{laloux1999noise, potters2005financial}. Furthermore, renewed attention has emerged in mathematical statistics, where advanced techniques grounded in free probability theory and deterministic equivalents have provided rigorous analytical tools for high-dimensional covariance estimation~\cite{yang2015robust, bun2017cleaning, Burda2022}.

In essence, RMT-based estimators improve covariance estimation by applying nonlinear shrinkage transformations to the eigenvalue spectrum while keeping the eigenvectors fixed~\cite{ledoit2004well, ledoit2011eigenvectors, ledoit2022power}. These estimators are known as Rotationally Invariant Estimators (RIEs). 
From this perspective, RMT-based estimators exploit the asymptotic properties of the eigenvalue distribution of the sample covariance matrix, which converges to a deterministic limit in the high-dimensional regime \( N/T = \mathcal{O}(1) \) \cite{bun2017cleaning, Burda2022}, corresponding to a system of \(N\) time series observed over \(T\) periods.

Complementary to this spectral approach, the effectiveness of filtering correlation matrices via hierarchical clustering has been extensively documented, particularly in the context of portfolio optimization in finance~\cite{tumminello2007hierarchically, tumminello2007spectral, tola2008cluster, pantaleo2011improved, bongiorno2021covariance, bongiorno2022reactive, garcia2023two, garcia2024random, garcia2025high}. At the same time, recent advances in machine learning have opened new avenues for high-dimensional covariance estimation, including deep learning–based methods for eigenvalue regularization within invariant estimator frameworks~\cite{bongiorno2025end} and reinforcement learning approaches to adaptive shrinkage estimation~\cite{mattera2023shrinkage}.

Despite these advances, the role of eigenvectors has remained largely secondary within most existing frameworks. In particular, many denoising techniques rely on unstructured assumptions about the population covariance matrix that fail to reproduce the stylized facts of complex systems, thereby substantially limiting their range of applicability. When structure is introduced, as in the works of~\cite{garcia2023two, garcia2025high}, both nested hierarchical covariance models and block-diagonal covariance structures give rise to eigenvalue spectra consistent with the so-called spiked covariance matrix model~\cite{johnstone2001distribution, donoho2018optimal}, namely covariance matrices characterized by a small number of large, isolated eigenvalues that are well separated from the bulk of the spectrum.

A key feature of spiked covariance models is top eigenvector inconsistency~\cite{donoho2018optimal}, meaning that the angle between sample eigenvectors and their corresponding population eigenvectors converges to a non-zero limit. This implies that filtering procedures based on Rotationally Invariant Estimators (RIEs), which retain sample eigenvectors, may fail to capture important structural properties of the population covariance matrix.

A further limitation of RIE methods arises for sample eigenvectors associated with small eigenvalues: these eigenvectors typically have components spread across the entire set of variables, producing orientations that differ substantially from the localized eigenvectors linked to the correlated dynamics of smaller, tightly connected groups of elements.   

The above observations motivate an alternative filtering procedure for spiked correlation matrices that combines machine learning with hierarchical clustering. In particular, we propose to explicitly exploit the information contained in the eigenvectors in order to improve covariance matrix estimation. Our approach performs a spectral decomposition of the covariance matrix, estimating the eigenvectors through a state-of-the-art deep neural network model, while the eigenvalues are obtained via the nonlinear shrinkage method based on Random Matrix Theory (RMT)~\cite{ledoit2020analytical}. In particular, we employ the Residual Neural Network (ResNet) architecture~\cite{he2016deep,he2020resnet}, which is well-suited for deep architectures as it mitigates computational challenges by introducing shortcut connections that facilitate efficient gradient propagation toward optimal solutions.

This proposal is benchmarked against two alternative covariance estimators: one entirely based on the ResNet architecture and another fully derived from Random Matrix Theory (RMT). Additionally, we include the two-step estimator introduced in~\cite{garcia2023two}, which incorporates the nested hierarchical organization of financial markets into advanced covariance estimation techniques. To the best of our knowledge, this is the first work to propose a hybrid estimation framework that combines the theoretical robustness of Random Matrix Theory with the representational capacity of deep neural networks applied to eigenvector information, specifically leveraging a ResNet architecture.

Since no analytical results are currently available regarding the denoising optimality of the proposed hybrid estimator—nor for the hierarchical clustering estimator or its two-step variant—we assess their performance through a comprehensive set of Monte Carlo experiments under a variety of complex population covariance structures. In particular, we consider a fully hierarchical covariance model \cite{garcia2025high} and a block-diagonal covariance model \cite{garcia2023two}. In addition, we introduce a novel population covariance model designed to reproduce key stylized facts of financial market interactions. This model generates a covariance matrix whose eigenvalue spectrum exhibits power-law scaling behavior in the scree plot, thereby capturing salient features of the multiscale dynamics observed in financial markets.

To validate the models in practice, we have considered a dataset composed of financial instruments known as cryptocurrencies. These assets are characterized by complex features that push traditional statistical methods to their limits. Specifically, cryptocurrencies do not follow a Gaussian distribution, exhibit heavy tails, abrupt jumps, asymmetry, and overall complex dynamics~\cite{Watorek2021Multiscale}.  Here, we carefully excluded pseudo-cryptocurrencies—i.e., coins that are merely replicas of fiat money—and retained only genuine blockchain projects or those that replicate exchange rates. Additionally, we pushed the estimators, particularly the deep learning ones, to their limits by training on a bull market period and testing on a bear market. This setup allows us to evaluate the model’s ability to handle structural changes and market regime shifts.

We find that both the nested hierarchical and power-law covariance models provide an adequate characterization of the complex interactions observed in cryptocurrency markets. Moreover, in both simulation studies and empirical cryptocurrency data, hybrid approaches—and, more broadly, deep learning–based methods—lead to statistically meaningful improvements in population covariance matrix estimation and, in turn, to enhanced out-of-sample financial performance in walk-forward analyses.
Although the covariance matrix is not the sole ingredient of an investment strategy, its accurate estimation plays a central role in controlling portfolio volatility across different market conditions, with implications that extend beyond the classical portfolio optimization paradigm~\cite{makridakis2024avoiding}.

Nevertheless, it is important to emphasize that the scope of the proposed approach is not restricted to financial systems. More generally, any high-dimensional system whose dependence structure can be modeled as a low-rank deformation of a Wishart ensemble may benefit from hybrid covariance estimators that explicitly incorporate eigenvector information. Such hybrid approaches are particularly relevant for systems that inherently exhibit complex nested, hierarchical, or multiscale structures, where purely eigenvalue-based denoising techniques may fail to capture essential geometric and structural properties encoded in the eigenvectors. From a statistical-physics perspective, this corresponds to exploiting information beyond spectral outliers—namely, the collective organization of eigenmodes—which plays a central role in inference problems exhibiting detectability phase transitions~\cite{johnstone2001distribution, bloemendal2013limits, feral2009largest}.

Section 2 presents the investment strategy used to allocate capital across the set of cryptocurrencies. In Section 3, we introduce the covariance matrix estimators derived from both RMT and deep learning (ResNet), as well as the proposed hybrid and two-step estimators. Section 4 details the proposed covariance matrix models used in the simulations, highlighting that the power-law model represents a novel contribution not previously discussed in the literature. Section 5 presents and discusses the results of the simulations, while Section 6 analyzes the findings related to the cryptocurrency dataset. Finally, Section 7 concludes the study and discusses potential avenues for future research.

\section{Asset allocation models}
\label{allocation_models}
\subsection{Portfolio Theory}
% Consider $p$ assets on $n$ trading days and denote by $s_{i,t}$ the price of asset $i=1,\dots,p$ at time $t=1,\dots,n$. The logarithmic return $r_{i,t}$ is defined as
%  \begin{equation}
%  \label{returns}
%  r_{i,t} = \log{(s_{i,t}/s_{i,t-1})}
%  \end{equation}
% The amount of money invested in the asset $i$ is known as the portfolio weight and is given by the vector
% \begin{equation}
%   \mathbf{w} = (w_1,\dots,w_p)^T,
%  \end{equation}
% A positive weight represents a \emph{long position} or ownership of an asset, whereas a negative weight indicates a \emph{short position}, meaning the investor sells a borrowed stock, expecting to buy it later at a lower
% price.

Consider $p$ assets observed over $n$ trading days, and let $s_{i,t}$ denote the price of asset $i = 1, \dots, p$ at time $t = 1, \dots, n$. The logarithmic return $r_{i,t}$ is defined as
\begin{equation}
\label{returns}
r_{i,t} = \log\left(\frac{s_{i,t}}{s_{i,t-1}}\right).
\end{equation}
The amount of capital invested in asset $i$ is represented by its portfolio weight, collected in the vector
\begin{equation}
\mathbf{w} = (w_1, \dots, w_p)^{\top}.
\end{equation}
A positive weight corresponds to a \emph{long position}, i.e., ownership of an asset, whereas a negative weight indicates a \emph{short position}, meaning that the investor sells a borrowed asset, expecting to repurchase it later at a lower price.

% Then, the portfolio return $\mathbf{R}$ is simply the dot product
% \begin{equation}
%     \mathbf{R} = \mathbf{w}^T \mathbf{r}
% \end{equation}
% Further, the expected return of the portfolio $\mathbf{\mu}$ is defined as
% \begin{equation}
% \mathbf{\mu}_R = \mathbb{E}(\mathbf{R}) = \mathbf{w}^T\mathbb{E}(\mathbf{r}) = \mathbf{w}^T\mathbf{\mu},
% \label{expected_profit}
% \end{equation}
% and the portfolio variance is expressed as a function of the population covariance matrix $\mathbf{\Sigma}$ of the returns
% \begin{equation}
%     \sigma^2_R = \mathbf{w}^T\mathbf{\Sigma w}.
% \end{equation}
% Thus, the volatility is computed as the standard deviation simply applying the square root 
% \begin{equation}
%     \sigma_R = \sqrt{\mathbf{w}^T\mathbf{\Sigma w}}.
%     \label{volatility}
% \end{equation}

Then, the portfolio return $\mathbf{R}$ is given by the dot product
\begin{equation}
    \mathbf{R} = \mathbf{w}^{\top} \mathbf{r}.
\end{equation}
The expected return of the portfolio, denoted by $\mu_R$, is defined as
\begin{equation}
    \mu_R = \mathbb{E}[\mathbf{R}] 
    = \mathbf{w}^{\top} \mathbb{E}[\mathbf{r}]
    = \mathbf{w}^{\top} \boldsymbol{\mu},
    \label{expected_profit}
\end{equation}
where $\boldsymbol{\mu}$ is the vector of expected returns for the individual assets.  

The portfolio variance is expressed as a quadratic form of the population covariance matrix $\boldsymbol{\Sigma}$ of the asset returns:
\begin{equation}
    \sigma_R^2 = \mathbf{w}^{\top} \boldsymbol{\Sigma} \mathbf{w}.
\end{equation}
Accordingly, the portfolio volatility is obtained as the square root of the variance:
\begin{equation}
    \sigma_R = \sqrt{\mathbf{w}^{\top} \boldsymbol{\Sigma} \mathbf{w}}.
    \label{volatility}
\end{equation}

\subsection{Minimum Variance Portfolio~(MVP)}

% The mean-variance allocation strategy of Markowitz~\cite{markowitz1952} proposes to solve the following quadratic optimization problem to minimize the portfolio risk at a given level of expected return~\cite{roncalli2013introduction}
% \begin{equation} 
% \underset{\mathbf{w}(\phi)\in\mathbb{R}^p}{max} \mathbf{w}^T \mathbf{\mu} - \frac{\phi}{2} \mathbf{w}^T \mathbf{\Sigma w}\quad\text{subject to}\quad \mathbf{1}^T\mathbf{w} = 1,
% \label{mean_variance}
% \end{equation}
% where $\phi$ is interpreted as the risk-aversion parameter.
% Hence, if $\phi=\infty$ the problem is transformed to
% \begin{equation} 
% \underset{\mathbf{w}(\infty)\in\mathbb{R}^p}{min} \frac{1}{2} \mathbf{w}^T\mathbf{\Sigma w}\quad\text{subject to}\quad \mathbf{1}^T\mathbf{w} = 1.
% \label{mv}
% \end{equation}

The mean--variance allocation strategy proposed by Markowitz~\cite{markowitz1952} seeks to solve the following quadratic optimization problem in order to minimize portfolio risk for a given level of expected return~\cite{roncalli2013introduction}:
\begin{equation} 
\label{mean_variance}
\underset{\mathbf{w}(\phi)\in\mathbb{R}^p}{\max} \;
\mathbf{w}^{\top} \boldsymbol{\mu}
- \frac{\phi}{2}\,\mathbf{w}^{\top} \boldsymbol{\Sigma} \mathbf{w}
\quad \text{subject to} \quad 
\mathbf{1}^{\top}\mathbf{w} = 1,
\end{equation}
where $\phi$ is interpreted as the risk-aversion parameter.  

When $\phi \to \infty$, the problem reduces to minimizing portfolio variance regardless of expected return:
\begin{equation}
\label{mv}
\underset{\mathbf{w}(\infty)\in\mathbb{R}^p}{\min} \;
\frac{1}{2}\,\mathbf{w}^{\top}\boldsymbol{\Sigma}\mathbf{w}
\quad \text{subject to} \quad
\mathbf{1}^{\top}\mathbf{w} = 1.
\end{equation}

% The reduced problem minimizes the volatility and is known as the Minimum Variance Portfolio~(MVP). 
% Further, it is possible to include the no short-selling restriction by considering the standard quadratic programming~(QP) problem. In this case, $\mathbf{w} \geq 0$, and a numerical solution can only be obtained~\cite{boyd2004convex}. This portfolio is also known as the long-only MVP because it does not allow short positions or negative weights. We will denote this portfolio as MVP+ in the subsequent analysis.

The reduced problem minimizes portfolio volatility and is referred to as the Minimum Variance Portfolio~(MVP). Moreover, it is possible to incorporate a no short-selling constraint by formulating the standard quadratic programming~(QP) problem. In this case, the portfolio weights satisfy $\mathbf{w} \geq 0$, and a numerical solution is required~\cite{boyd2004convex}. This constrained portfolio is also known as the long-only MVP, as it does not permit short positions or negative weights. In the following analysis, we will denote this portfolio as MVP$^+$.

\section{Covariance estimators}
\label{estimators}
\subsection{Random matrix denoising}

% A \emph{naive estimator} of the population covariance matrix $\mathbf{\Sigma}$ given the empirical or sample covariance matrix $\mathbf{S}$ is given by
% \begin{equation}
%     \mathbf{\Xi}^{naive} = \mathbf{S},
% \end{equation} 
% which is a unbiased estimator of $\mathbf{\Sigma}$ whenever the number of variables $p$ is fixed and the number of observations $n\rightarrow\infty$\cite{johnson2002applied}. 

% A non-linear shrinkage formula minimizing the Frobenious loss functions has been proposed by Ledoit and Peche~\cite{ledoit2011eigenvectors} from Random Matrix Theory~(RMT) principles~\cite{potters2020first}. The covariance matrix in high dimensions is estimated by the following formula
% \begin{eqnarray}
% \mathbf{\Xi}^{LP} &=& \sum_{k=1}^{p} \xi_k^{LP} v_k v_k^T,\quad
%     \text{where}\\
% \xi^{LP}_k  &=&  \lim_{\epsilon\rightarrow 0^{+}}\frac{\lambda_k}{|1-q+q\lambda_k G_S(\lambda_k-i\epsilon)|^2},
% \label{LedoitPeche2011}    
% \end{eqnarray}
% where $\lambda_k, v_k$  are the eigenvalue and eigenvector tuple of $\mathbf{S}$, and $G_S$ is the Stieltjes transform of $\mathbf{S}$. 

A \emph{naive estimator} of the population covariance matrix $\boldsymbol{\Sigma}$, given the empirical or sample covariance matrix $\mathbf{S}$, is simply
\begin{equation}
    \boldsymbol{\Xi}^{\text{naive}} = \mathbf{S},
\end{equation} 
which is an unbiased estimator of $\boldsymbol{\Sigma}$ when the number of variables $p$ is fixed and the number of observations $n \to \infty$~\cite{johnson2002applied}.  

A non-linear shrinkage formula that minimizes the Frobenius loss has been proposed by Ledoit and Péché~\cite{ledoit2011eigenvectors}, based on principles from RMT~\cite{potters2020first}. In high-dimensional settings, the covariance matrix is estimated as
\begin{align}
\boldsymbol{\Xi}^{\text{LP}} &= \sum_{k=1}^{p} \xi_k^{\text{LP}}\, \mathbf{v}_k \mathbf{v}_k^{\top}, \quad
\text{where} \\
\xi^{\text{LP}}_k &= \lim_{\epsilon \to 0^+} 
\frac{\lambda_k}{\big|1 - q + q \lambda_k G_S(\lambda_k - i\epsilon)\big|^2},
\label{LedoitPeche2011}    
\end{align}
and $(\lambda_k, \mathbf{v}_k)$ denote the eigenvalue--eigenvector pairs of $\mathbf{S}$, and $G_S$ is the Stieltjes transform of $\mathbf{S}$.

\subsection{Residual Neural Network denoising}

An estimator based on machine learning is proposed, employing the Residual Neural Network (ResNet) architecture~\cite{he2016deep}. Starting from the sample covariance matrix $\mathbf{S}$, the ResNet applies repeated residual updates of the form
\begin{equation}
\mathbf{S} \;\mapsto\; \mathbf{S} + \mathcal{F}(\mathbf{S}),
\end{equation}
where $\mathcal{F}$ denotes a generic neural network module. In the present implementation, $\mathcal{F}$ is constructed using Convolutional Neural Networks (CNNs)~\cite{brownlee2018deep}, which are well suited for processing grid-structured inputs and for exploiting local spatial dependencies within the input representation.
ResNet improves CNNs by introducing a novel neural architecture that incorporates \emph{skip connections}~\cite{nielsen2015neural}, allowing information from one layer to propagate to non-contiguous layers. Specifically, ResNet contains connections between layer $i$ and layer $(i + r)$ for $r > 1$, where $i$ is an arbitrary layer and $r$ denotes the skip length. 

\begin{figure}[hbtp]
    \centering
    \begin{subfigure}[b]{0.45\textwidth}
    \includegraphics[width=0.6\linewidth]{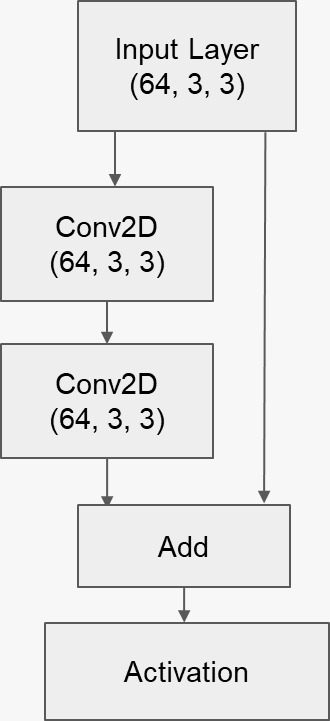}
        \caption{}
    \end{subfigure}
    \begin{subfigure}[b]{0.45\textwidth}
     \includegraphics[width=0.95\linewidth]{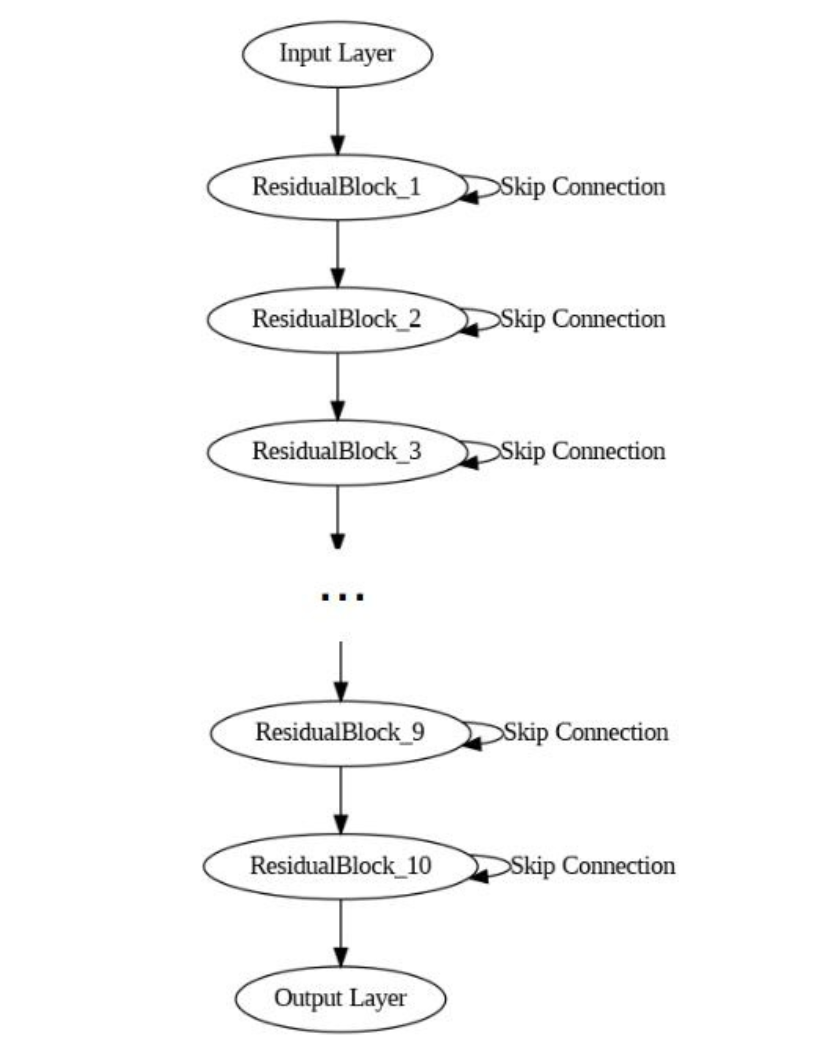}
     \caption{}
    \end{subfigure}
    \caption{Network architecture. (a) Residual block with $r=2$. (b) ResNet with 10 basic residual blocks.}
    \label{architecture}
\end{figure}

The CNN learning process operates as follows: a \emph{filter} is applied to the input to detect local patterns, producing a corresponding \emph{feature map}. An activation function is then applied element-wise after each layer to introduce nonlinearity, regulate signal propagation, and constrain the output to a desired range of values. Figure~\ref{architecture}(a) illustrates the architecture of the residual block used for the covariance matrix denoising, setting $r = 2$.  

Here, the residual module is constructed with an initial convolutional layer comprising 64 filters of size $3 \times 3$, producing 64 feature maps followed by a rectified linear unit (ReLU) activation. This base representation is subsequently refined through two stacked two-dimensional convolutional layers (Conv2D) to extract higher-level features. The first Conv2D layer employs a ReLU activation function, whereas the second uses a linear activation function.

The \texttt{add} block implements the skip connection by combining the module input (the residual) with the output of the stacked Conv2D layers. In this formulation, the network learns only the residual correction rather than the full mapping, thereby mitigating the vanishing gradient problem. More generally, skip connections facilitate effective gradient propagation, allowing the optimization procedure to adaptively regulate the degree of nonlinearity applied to a given input.

Finally, the residual module applies a ReLU activation at its output. The complete network architecture, illustrated in Fig.~\ref{architecture}(b), is obtained by stacking ten such residual blocks. Using this ResNet architecture, together with appropriate symmetrization and a positive semidefinite (PSD) projection, the network learns to attenuate noise in the empirical covariance matrix. We denote the resulting estimator by $\boldsymbol{\Xi}^{\mathrm{CNN}}$, highlighting the CNN-based structure of its internal layers.

We further propose a hybrid approach that leverages both the learning capabilities of the ResNet model and the theoretical properties of high-dimensional covariance estimation. The core idea is first to perform a spectral decomposition of the empirical covariance matrix:
\begin{equation}
    \mathbf{S} = \mathbf{V} \mathbf{\Lambda} \mathbf{V}^T
\end{equation}

Here, $\mathbf{V}$ denotes the $p \times p$ matrix whose $k$-th column is the eigenvector $\mathbf{v}_k$ of $\mathbf{S}$, and $\mathbf{\Lambda}$ is the diagonal matrix containing the corresponding eigenvalues $\lambda$s.  
The hybrid estimator applies the ResNet learning model to the eigenvectors (without symmetrization or PSD transformation), while the eigenvalues are processed using the nonlinear shrinkage RMT-based formula in Eq.~\eqref{LedoitPeche2011}. Formally, the hybrid estimator is defined as
\begin{equation}
    \boldsymbol{\Xi}^{H} = \boldsymbol{\Xi}^{\text{CNN}}(\mathbf{V}) \, \boldsymbol{\Xi}^{\text{LP}}(\mathbf{\Lambda}) \, \boldsymbol{\Xi}^{\text{CNN}}(\mathbf{V}^{\top}),
\end{equation}
where $\boldsymbol{\Xi}^{\text{CNN}}(\mathbf{V})$ and $\boldsymbol{\Xi}^{\text{LP}}(\mathbf{\Lambda})$ denote the application of the operators $\boldsymbol{\Xi}^{\text{CNN}}$ and $\boldsymbol{\Xi}^{\text{LP}}$ to the matrix of eigenvectors and the diagonal matrix of eigenvalues, respectively.
This hybrid approach combines the data-driven flexibility of ResNet to capture complex dependencies in the eigenvectors, while exploiting the theoretical guarantees of RMT for consistent eigenvalue estimation in high-dimensional settings.

\subsection{Hierarchical clustering denoising}
% A novel approach to estimate the covariance matrix is proposed in \cite{tumminello2007kullback} using a hierarchical clustering algorithm. 
% The procedures consist of transforming the empirical covariance matrix $\mathbf{S}$ into a correlation matrix $\mathbf{C}$.
% Then, the transformation $\mathbf{D}=\mathbf{11}^T-\mathbf{C}$ is applied, which also satisfies the axioms of a distance measure, being  $\mathbf{1}$ a vector of ones of dimension $p$.
% Then, a dendrogram of $\mathbf{D}$ was constructed via the Average Linkage Clustering Analysis~(ALCA)\cite{johnson2002applied} and the distance $\rho$ between clusters at each hierarchical level was computed. Hence, we can obtain a dissimilarity matrix  $\mathbf{D}(\rho)$ as a function of $\rho$, and the filtered covariance matrix is retrieved by the inverse transformation $\mathbf{\Xi}^{ALCA}=\mathbf{H}^{1/2}(\mathbf{11}^T-\mathbf{D}(\rho))\mathbf{H}^{1/2}$.

A novel approach to covariance matrix estimation was proposed by Tumminello et al.~\cite{tumminello2007kullback} using a hierarchical clustering algorithm. The procedure begins by transforming the empirical covariance matrix $\mathbf{S}$ into the corresponding correlation matrix $\mathbf{C}$.  
Next, the transformation
\begin{equation}
    \mathbf{D} = \mathbf{11}^{\top} - \mathbf{C}
\end{equation}
is applied, which satisfies the axioms of a distance measure, where $\mathbf{1}$ is a $p$-dimensional vector of ones. A dendrogram is then constructed from $\mathbf{D}$ using Average Linkage Clustering Analysis (ALCA)~\cite{johnson2002applied}, and the distance $\rho$ between clusters at each hierarchical level is computed.  
This procedure yields a dissimilarity matrix $\mathbf{D}(\rho)$ as a function of $\rho$, from which the filtered covariance matrix is recovered via the inverse transformation:
\begin{equation}
    \boldsymbol{\Xi}^{\text{ALCA}} = \mathbf{H}^{1/2} \, (\mathbf{11}^{\top} - \mathbf{D}(\rho)) \, \mathbf{H}^{1/2},
\end{equation}
where $\mathbf{H}$ is the diagonal matrix of variances used to rescale the correlation matrix back to the original covariance scale.

A state-of-the-art estimator designed to address both the heterogeneous structure of financial markets and the challenges of high-dimensional settings was proposed in~\cite{garcia2023two}. The core idea is to first apply our proposed single-step estimators, followed by a hierarchical filtering step. We consider the following combinations:

\begin{align}
    \boldsymbol{\Xi}^{2S(\text{LP})} &:= \boldsymbol{\Xi}^{\text{ALCA}}\big(\boldsymbol{\Xi}^{\text{LP}}\big),\\
    \boldsymbol{\Xi}^{2S(\text{CNN})} &:= \boldsymbol{\Xi}^{\text{ALCA}}\big(\boldsymbol{\Xi}^{\text{CNN}}\big),\\
    \boldsymbol{\Xi}^{2S(\text{H})} &:= \boldsymbol{\Xi}^{\text{ALCA}}\big(\boldsymbol{\Xi}^{H}\big).
\end{align}

% The mathematical justification for this estimator is discussed in \cite{garcia2023two} and is based on the following arguments: (i) the application of an RMT estimator in the first step reduces the noise or estimation error of the largest eigenvalues, and (ii) applying a filtering process as a second step via hierarchical clustering can mitigate the inconsistency in the top eigenvectors, which is inevitably associated with the RMT covariance estimators \cite{donoho2018optimal}.
%Here the explanations is extended to the new proposals $\boldsymbol{\Xi}^{\text{CNN}}$ and $\boldsymbol{\Xi}^{H}$, interpreting the second step as a regularization step to avoid overfitting.

The mathematical justification for this two-step estimator is discussed in~\cite{garcia2023two} and is based on the following arguments: 
(i) applying an RMT-based estimator in the first step reduces the noise and estimation error associated with the largest eigenvalues, and 
(ii) applying a subsequent filtering step via hierarchical clustering mitigates the inconsistency of the top eigenvectors, which is inherently present in RMT covariance estimators~\cite{donoho2018optimal}.
Here, this reasoning is extended to the newly proposed estimators $\boldsymbol{\Xi}^{\text{CNN}}$ and $\boldsymbol{\Xi}^{H}$, where the second step can be interpreted as a regularization procedure to avoid overfitting.

\section{Covariance models}
% To test our covariance matrix estimators, we consider the following data-generating process
% \begin{eqnarray}
%     \mathbf{Y} = \sqrt{\mathbf{\Sigma}}\mathbf{X}\\
%     \label{dgp}
% \end{eqnarray}
% where $\mathbf{X}$ is a $p\times n$ random matrix, with each entry following a standard Gaussian distribution, and $\mathbf{\Sigma}$ is a $p\times p$ population covariance matrix.
% In this way, given a model for $\mathbf{\Sigma}$, we can obtain the sample covariance realization
% \begin{eqnarray}
%     \mathbf{S} = \frac{1}{n} \mathbf{Y} \mathbf{Y}^T =  \frac{1}{n} \sqrt{\mathbf{\Sigma}} \mathbf{X}\mathbf{X}^T\sqrt{\mathbf{\Sigma}}
%     \label{noise_model}
% \end{eqnarray}
% In this work, we examine the following covariance population models of $\mathbf{\Sigma}$

To evaluate the performance of the proposed covariance matrix estimators, we consider the following data-generating process:
\begin{equation}
    \mathbf{Y} = \sqrt{\boldsymbol{\Sigma}}\,\mathbf{X},
    \label{dgp}
\end{equation}
where $\mathbf{X}$ is a $p \times n$ random matrix whose entries are independent and identically distributed according to a standard Gaussian distribution, and $\boldsymbol{\Sigma}$ is a $p \times p$ population covariance matrix. 
Given a model specification for $\boldsymbol{\Sigma}$, we can obtain the sample covariance realization as
\begin{equation}
    \mathbf{S} = \frac{1}{n}\mathbf{Y}\mathbf{Y}^T = \frac{1}{n}\sqrt{\boldsymbol{\Sigma}}\,\mathbf{X}\mathbf{X}^T\sqrt{\boldsymbol{\Sigma}},
    \label{noise_model}
\end{equation}
which represents the noisy empirical counterpart of the true covariance matrix. 
In this study, we analyze the following population models for $\boldsymbol{\Sigma}$.

\begin{figure}[hbtp]
    \centering
    \begin{subfigure}[b]{0.65\textwidth}
        \includegraphics[scale=0.45]{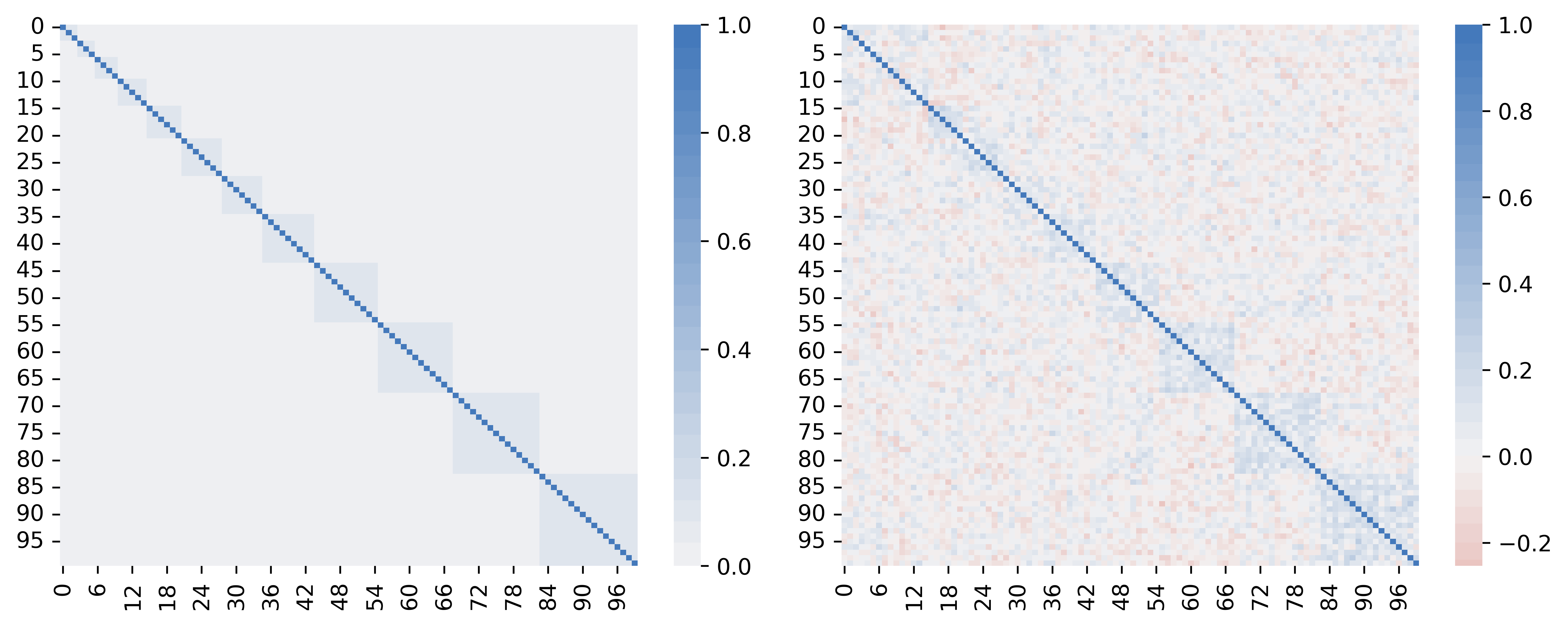}
        \caption{}
    \end{subfigure}\\
    \begin{subfigure}[b]{0.65\textwidth}
        \includegraphics[scale=0.45]{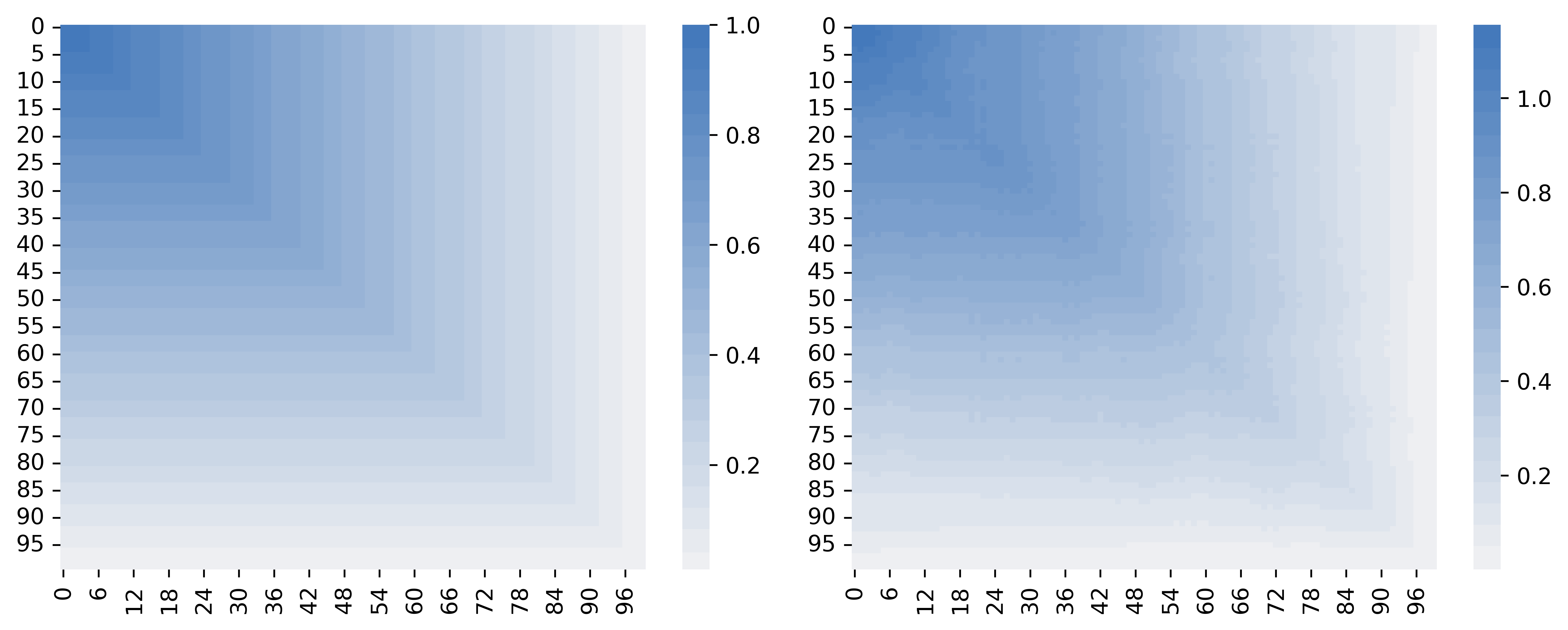}
        \caption{}
    \end{subfigure}\\
    \begin{subfigure}[b]{0.65\textwidth}
       \includegraphics[scale=0.45]{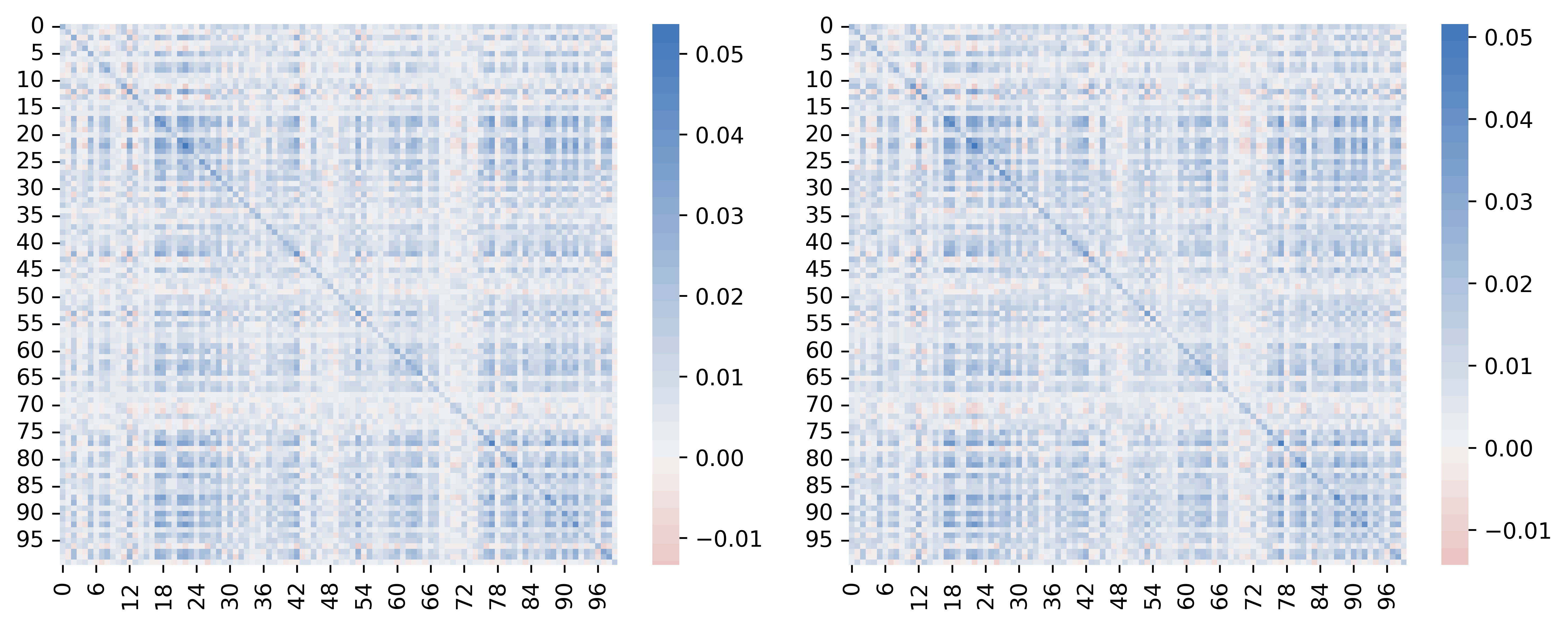}
        \caption{}
    \end{subfigure}
    \caption{Covariance models. (a) Block diagonal correlation model. (b) Nested hierarchical model. (c) Power-law model. Left: population covariances. Right: sample realization.}
    \label{fig_cov_models}
\end{figure}

\begin{itemize}
% \item[(1)] A block diagonal correlation model with the structure:
%     \begin{equation}
%     \mathbf{\Sigma} = \mathbf{LL}^T
%     \label{covariance_structere}
%     \end{equation}

% where the matrix $\mathbf{L}$ is a rectangular matrix of dimension $p\times k$, being $k$ the numbers of blocks. In particular, we consider $p=100, k=12$ with blocks of heterogeneous sizes equal to $[3,3,4,5,6,7,7,9,11,13,15,17]$, but same intensities $\gamma=0.3$. To have a proper correlation matrix, the diagonal entries are filled with ones. Figure~\ref{fig_cov_models}(a) shows the population model for these specifications (left) and a finite sample realization with $n=200$~(right). 
% This block-independent and homogeneous structure model has been previously studied in~\cite {garcia2023two}.

\item[(1)] A block-diagonal correlation structure defined as
\begin{equation}
    \boldsymbol{\Sigma} = \mathbf{L}\mathbf{L}^T,
    \label{covariance_structure}
\end{equation}
where $\mathbf{L}$ is a rectangular matrix of dimension $p \times k$, with $k$ denoting the number of blocks. 
In particular, we set $p = 100$ and $k = 12$, with heterogeneous block sizes given by $[3,3,4,5,6,7,7,9,11,13,15,17]$, but with equal intra-block correlation intensity $\gamma = 0.3$. 
To ensure a proper correlation matrix, the diagonal entries are set to one. 
Figure~\ref{fig_cov_models}(a) displays the population model under these specifications (left) and a finite-sample realization with $n = 200$ (right). 
This block-independent and homogeneous structure model has been previously examined in~\cite{garcia2023two}.

% \item[(2)] A completely nested hierarchical covariance model, where using the same general structure of eq.~\ref{covariance_structure}, the heterogeneity is integrated through the matrix $\mathbf{L}$ of dimension $p\times p$ is given by
%     \begin{equation}
%     \mathbf{L} = \begin{pmatrix}
%         \gamma & \gamma & \dots & \gamma & \gamma \\
%         \gamma & \gamma & \dots &  \gamma & 0 \\
%         \vdots &  \vdots & \ddots & \vdots & \vdots \\
%         \gamma &  \gamma & \dots & 0 & 0 \\
%         \gamma &  0 & \dots & 0 & 0 \\
%     \end{pmatrix}
% \end{equation}
% Here, we have set $\gamma = 0.1$. Figure~\ref{fig_cov_models}(b) shows the population model for these specifications~(left) with $p=100$ and a finite sample realization with $n=200$~(right). This system has been proposed in~\cite{garcia2025high} as a model to characterize the complex interactions of the financial markets. In particular, the population eigenvalues of this system are given by the solution of a tridiagonal symmetric matrix and have deep connections with Fibonacci and Lucas numbers~\cite{cahill2004fibonacci}.

\item[(2)] A completely nested hierarchical covariance model, where using the same general structure of eq.~\ref{covariance_structure}, the heterogeneity is integrated through the matrix $\mathbf{L}$ of dimension $p\times p$ is given by
\begin{equation}
    \mathbf{L} = 
    \begin{pmatrix}
        \gamma & \gamma & \dots & \gamma & \gamma \\
        \gamma & \gamma & \dots & \gamma & 0 \\
        \vdots & \vdots & \ddots & \vdots & \vdots \\
        \gamma & \gamma & \dots & 0 & 0 \\
        \gamma & 0 & \dots & 0 & 0 \\
    \end{pmatrix}.
\end{equation}
Here, we set $\gamma = 0.1$. 
Figure~\ref{fig_cov_models}(b) presents the population model under these specifications (left) with $p = 100$, and a finite-sample realization with $n = 200$ (right). 
This system was originally proposed in~\cite{garcia2025high} as a model to characterize the complex interaction structure of financial markets. 
In particular, the population eigenvalues of this system correspond to the solutions of a symmetric tridiagonal matrix and exhibit deep mathematical connections with Fibonacci and Lucas numbers~\cite{cahill2004fibonacci}.

% \item[(3)] A power-law model of the form
%     \begin{equation}
%     \mathbf{\Sigma} = \mathbf{O}\mathbf{\Lambda} \mathbf{O}^T
%     \label{covariance_powerlaw}
%     \end{equation}
% where $\mathbf{O}$ is an orthogonal random matrix of dimension $p\times p$, and $\mathbf{\Lambda}$ is a diagonal matrix of the same dimension with entries:
% \begin{equation}
%     \lambda_i = i^{-\alpha},\quad i=1,\dots,p,
% \end{equation}
% representing the eigenvalues of $\mathbf{\Sigma}$.  By construction, these eigenvalues follow a power law behavior. Figure~\ref{fig_cov_models}(b) shows the population model for these specifications~(left) considering $p=100$, $\alpha=1.5$, and a finite sample realization with $n=200$~(right). As far as we know, this model is presented for the first time as a covariance matrix representation of power-law interactions. 

\item[(3)] A power-law model of the form
\begin{equation}
    \mathbf{\Sigma} = \mathbf{O}\mathbf{\Lambda}\mathbf{O}^T,
    \label{covariance_powerlaw}
\end{equation}
where $\mathbf{O}$ is a random orthogonal matrix of dimension $p \times p$, and $\mathbf{\Lambda}$ is a diagonal matrix of the same dimension with entries
\begin{equation}
    \lambda_i = i^{-\alpha}, \quad i = 1, \dots, p,
\end{equation}
representing the eigenvalues of $\mathbf{\Sigma}$. 
By construction, these eigenvalues follow a power-law decay. 
Figure~\ref{fig_cov_models}(c) displays the population model under these specifications (left), considering $p = 100$ and $\alpha = 1.5$, along with a finite-sample realization for $n = 200$ (right). 
To the best of our knowledge, this formulation is presented here for the first time as a covariance matrix representation of power-law interactions.
\end{itemize}

% Figure~\ref{fig_screeplot} shows the log-log screeplot of eigenvalues of each model.
% We can see that model 1 shows $k=12$ non-degenerate eigenvalues, whose magnitudes are associated with the size of each of the $k$ blocks (see~\cite{garcia2023two}).
% Model 2 has the intriguing characteristic that their scree plot of eigenvalues follows approximately a power law. Due to this characteristic, in~\cite{garcia2025high} it has been conjectured that it can capture the stylized fact of the complex interactions of the financial markets.
% In the case of model 3, the power-law behavior is given by construction with a slope of $\alpha=1.5$. Then, it is a one-step further proposal to characterize the complex interactions of financial markets through their covariance relationships.

\begin{figure}[hbtp]
    \centering
    \begin{subfigure}[b]{0.45\textwidth}
       \includegraphics[scale=0.25]{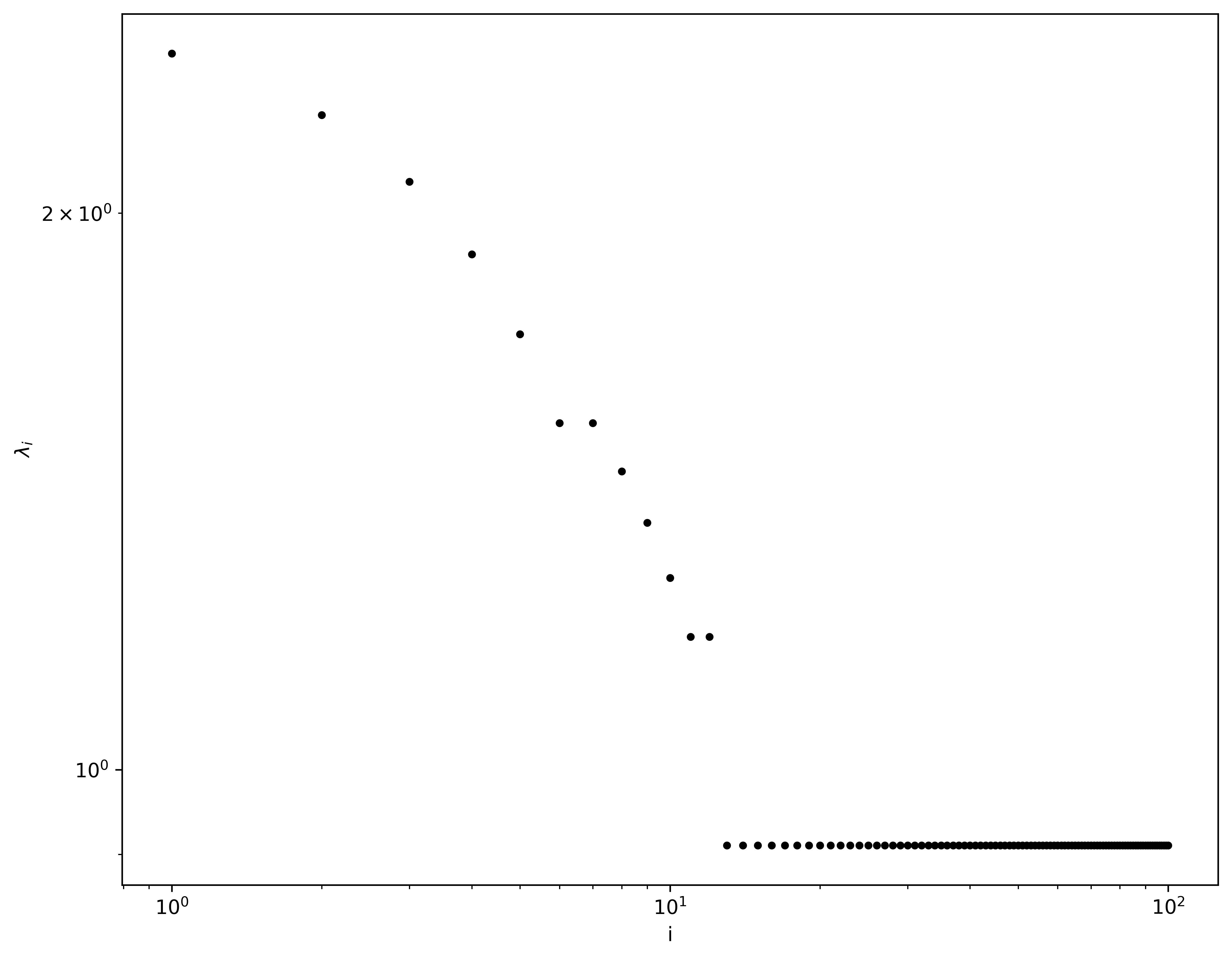}
        \caption{}
    \end{subfigure}\\
    \begin{subfigure}[b]{0.45\textwidth}
       \includegraphics[scale=0.25]{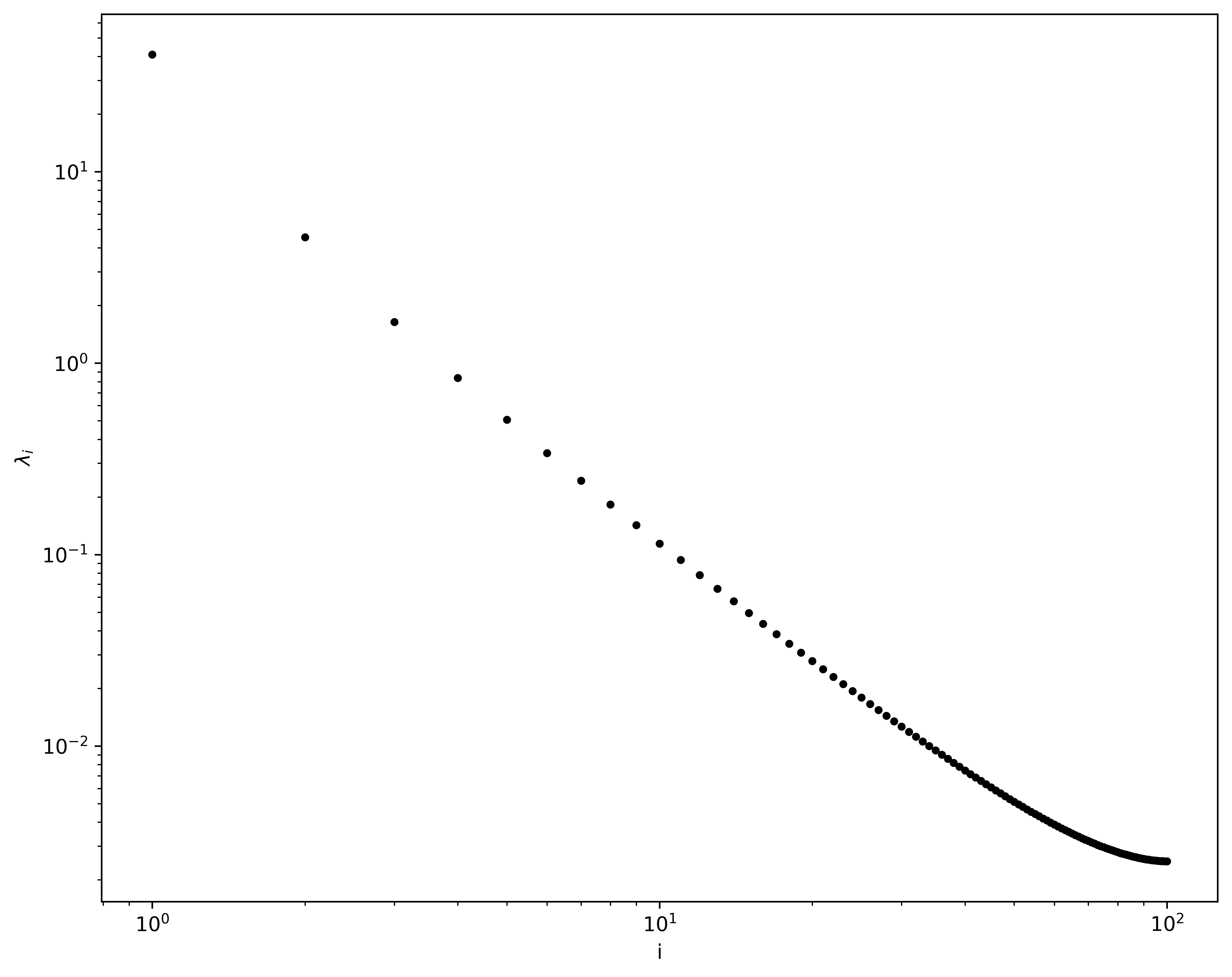}
        \caption{}
    \end{subfigure}\\
    \begin{subfigure}[b]{0.45\textwidth}
       \includegraphics[scale=0.25]{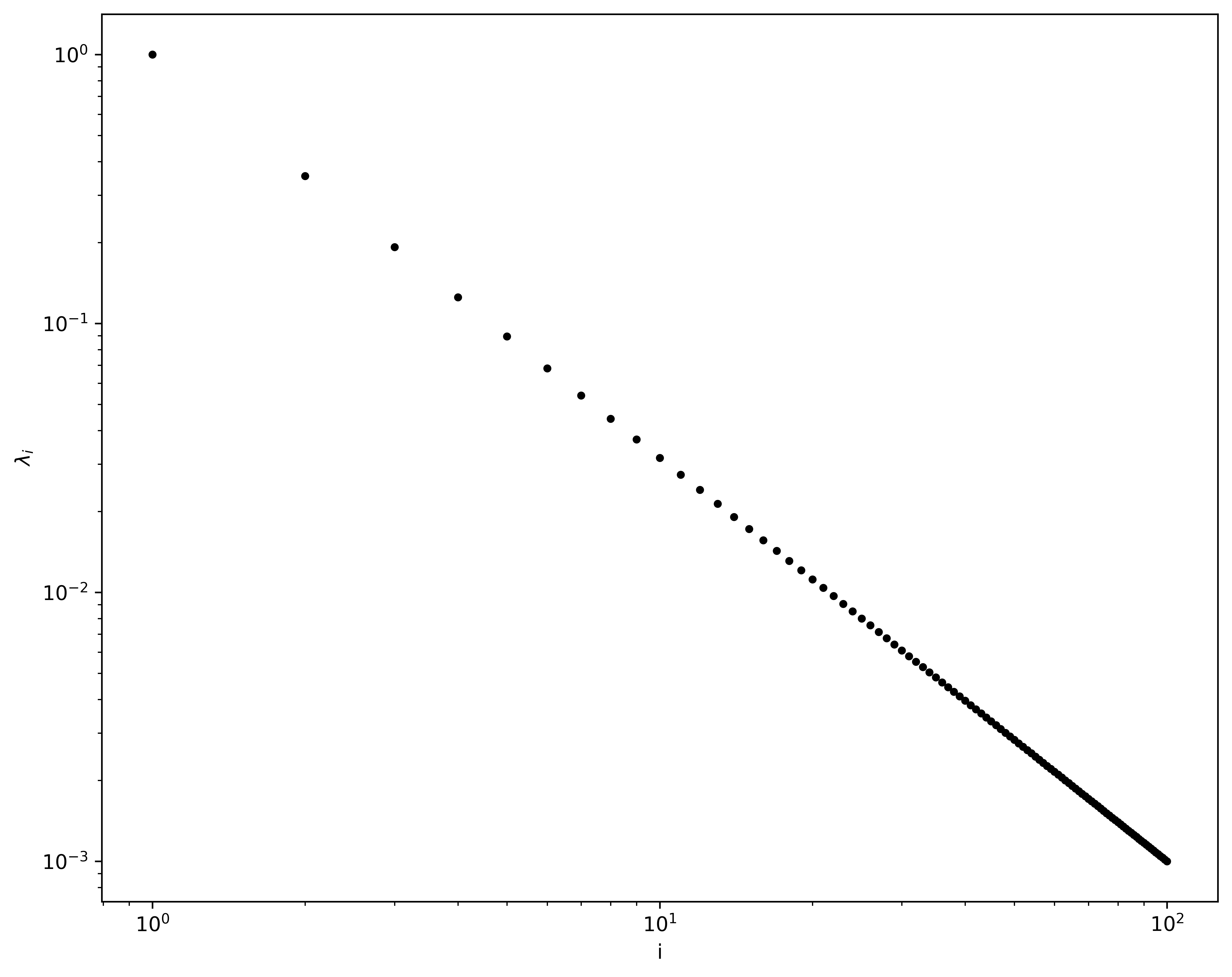}
        \caption{}
    \end{subfigure}
    \caption{Scree plot of eigenvalues for (a) model 1, (c) model 2, and (e) model 3.}
    \label{fig_screeplot}
\end{figure}

Figure~\ref{fig_screeplot} presents the log-log scree plot of the eigenvalues for each model. 
Model~1 exhibits $k = 12$ non-degenerate eigenvalues, whose magnitudes correspond to the size of each of the $k$ blocks (see~\cite{garcia2023two}). 
Model~2 displays the intriguing property that its eigenvalue spectrum approximately follows a power-law decay. 
Due to this property, it has been conjectured in~\cite{garcia2025high} that it can capture the stylized facts underlying the complex interactions of financial markets.
In the case of Model 3, the power-law behavior is imposed by construction with a slope of $\alpha = 1.5$, providing a further step toward characterizing the complex interactions of financial markets through their covariance structure.

% Figure~\ref{fig_dendrogram} shows the associated dendrogram of each covariance model following the same methodology of the hierarchical estimator but considering a single linkage variant~(SLCA). Model 1 shows a homogeneous hierarchical structure that recovers the $k=12$ blocks of the population model. By its part, model 2 shows hierarchies of increasing size as modeled in the population model. Finally, the dendrogram associated with model 3 also shows an increasing size of nested hierarchies. Nevertheless, the block structure is heterogeneous due to the inherited random signatures of the orthogonal random eigenvectors~(see eq.~\ref{covariance_powerlaw}).

\begin{figure}[hbtp]
    \centering
    \begin{subfigure}[b]{0.45\textwidth}
        \includegraphics[scale=0.25]{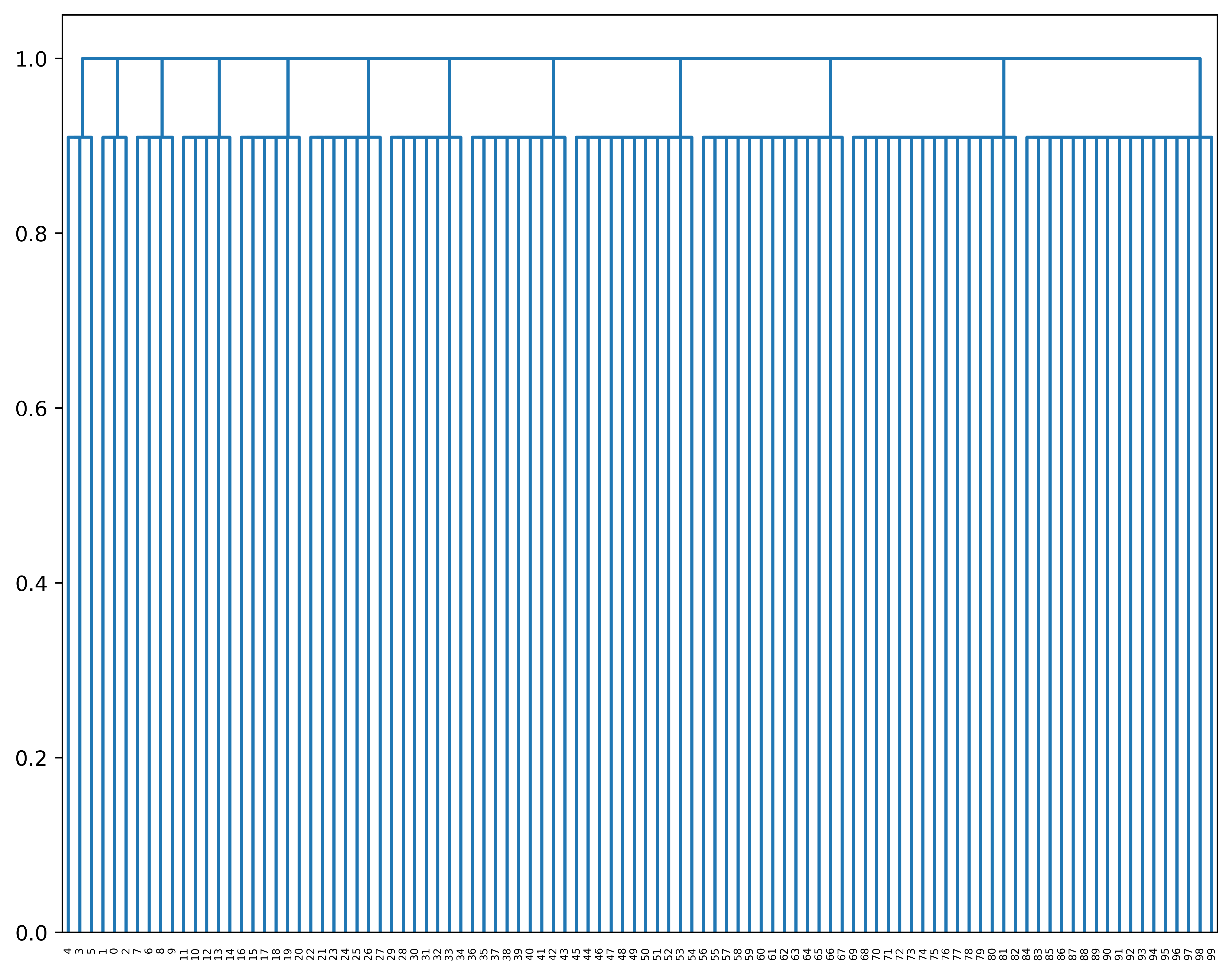}
        \caption{}
    \end{subfigure}\\
    \begin{subfigure}[b]{0.45\textwidth}
        \includegraphics[scale=0.25]{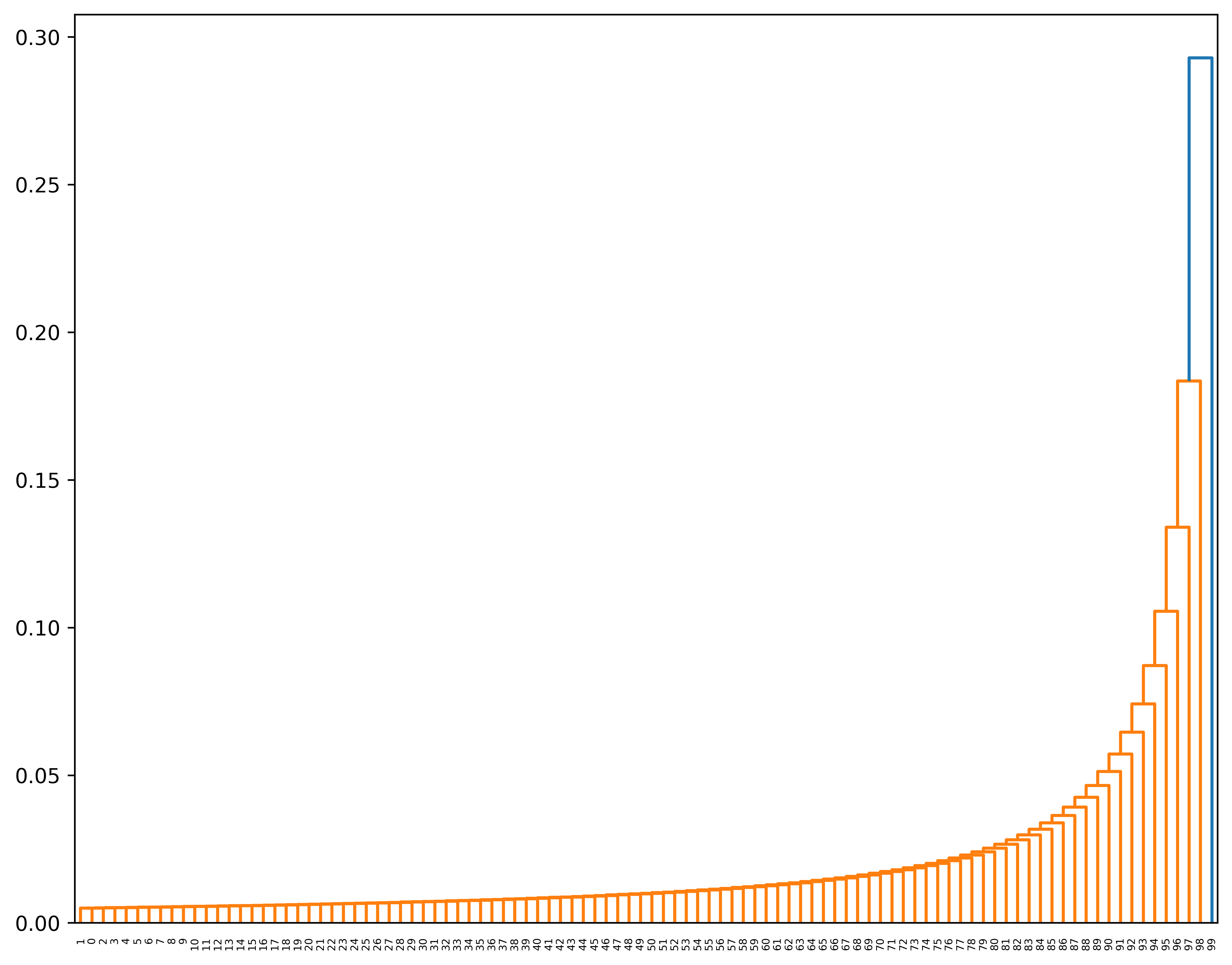}
        \caption{}
    \end{subfigure}\\
    \begin{subfigure}[b]{0.45\textwidth}
        \includegraphics[scale=0.25]{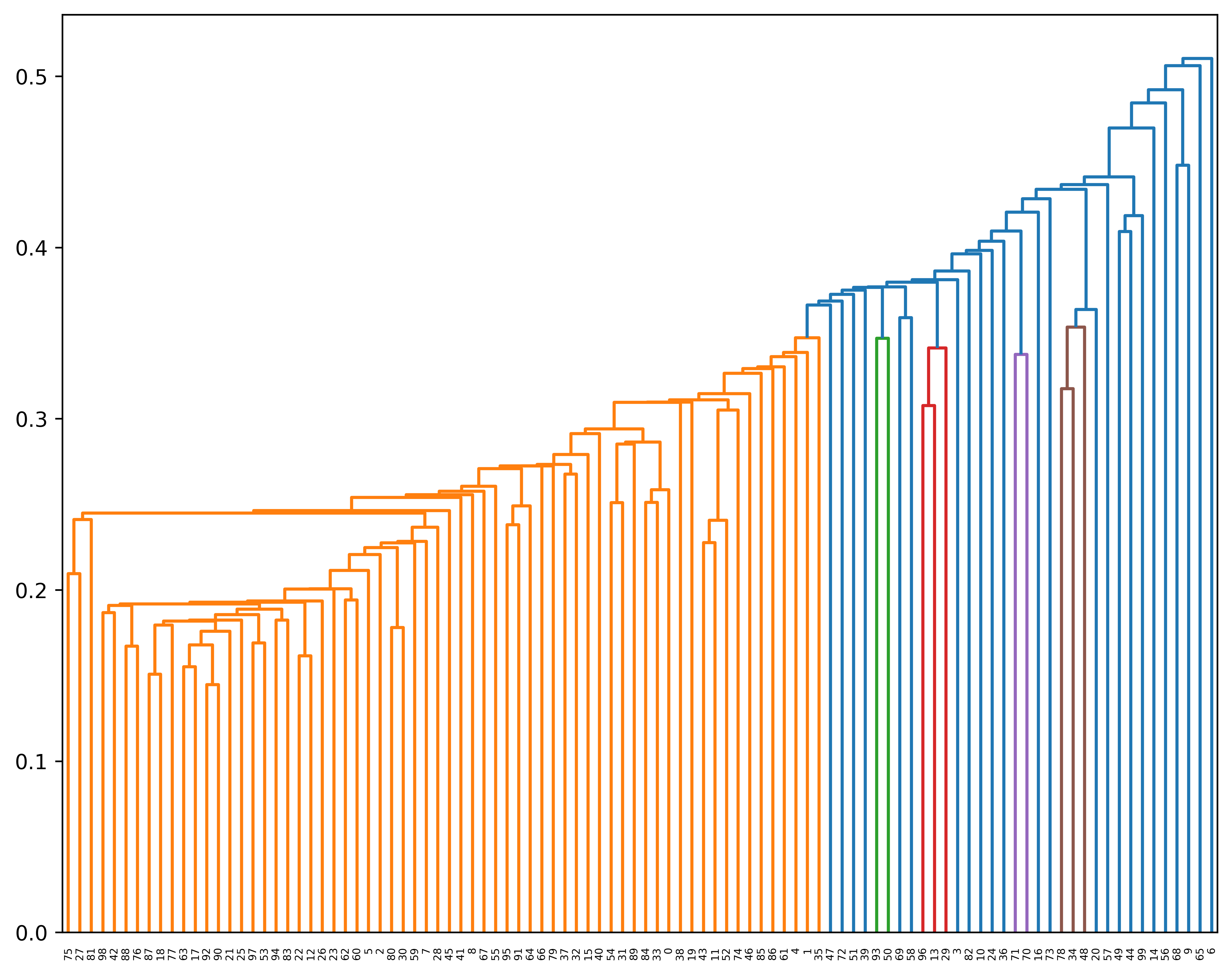}
        \caption{}
    \end{subfigure}\\
    \caption{Associated dendrogram of (a) model 1, (b) model 2, and (c) model 3. For illustrative purposes, the descendant links below a cluster node $k$ are equally colored if $k$ is the first node below the cut threshold $t(=0.7)$.}
    \label{fig_dendrogram}
\end{figure}

Figure~\ref{fig_dendrogram} displays the dendrograms associated with each covariance model, constructed using the same methodology as the hierarchical estimator but employing the Single Linkage Clustering Algorithm~(SLCA). 
Model~1 exhibits a homogeneous hierarchical structure that successfully recovers the $k = 12$ blocks of the population model. 
In contrast, Model~2 reveals hierarchies of increasing size, consistent with the structure imposed in the population model. 
Finally, the dendrogram corresponding to Model~3 also displays an increasing degree of nested hierarchies; however, its block structure is heterogeneous due to the random signatures inherited from the orthogonal eigenvectors (see Eq.~\ref{covariance_powerlaw}).

\section{Simulations}
% We perform a Monte Carlo simulation considering $m=1000$ sample covariance realizations of each of the three covariance population models. The sample covariance matrices are generated by finite data matrices of dimensions $p=100$, $n=100$. Thus, we filter out the noise of each sample by applying each of the covariance matrix estimators presented in section~\ref{estimators}.
% Then, we compute the Frobenious~($F$) and Minimum Variance~($MV$) loss functions between the filtered $\mathbf{\Xi}$ and population covariance model $\mathbf{\Sigma}$, which are given by the following expressions:
% \begin{eqnarray}
%     F(\mathbf{\Xi},\mathbf{\Sigma}) = \frac{1}{p}Tr[(\mathbf{\Xi}-\mathbf{\Sigma})(\mathbf{\Xi}-\mathbf{\Sigma})^T],\\
%     MV(\mathbf{\Xi},\mathbf{\Sigma}) = \frac{Tr(\mathbf{\Sigma}^{-1} \mathbf{\Xi} \mathbf{\Sigma}^{-1})/p}{[Tr(\mathbf{\Sigma}^{-1})/p]^2} - \frac{1}{Tr(\mathbf{\Xi}^{-1})/p}
% \end{eqnarray}

We conduct a Monte Carlo simulation consisting of $m = 1000$ realizations of the sample covariance matrix for each of the three population covariance models. 
The sample covariance matrices are generated from finite data matrices of dimensions $p = 100$ and $n = 200$. 
For each realization, the noise is filtered by applying the covariance matrix estimators introduced in Section~\ref{estimators}. 
Subsequently, we evaluate the performance of the estimators using the Frobenius~($F$) and Minimum Variance~($MV$) loss functions, computed between the filtered matrix $\mathbf{\Xi}$ and the true population covariance matrix $\mathbf{\Sigma}$. 
These loss functions are defined as follows:
\begin{eqnarray}
    F(\mathbf{\Xi}, \mathbf{\Sigma}) &=& \frac{1}{p} \operatorname{Tr}\!\left[(\mathbf{\Xi} - \mathbf{\Sigma})(\mathbf{\Xi} - \mathbf{\Sigma})^{T}\right],\\
    MV(\mathbf{\Xi}, \mathbf{\Sigma}) &=& 
    \frac{\operatorname{Tr}(\mathbf{\Sigma}^{-1} \mathbf{\Xi} \mathbf{\Sigma}^{-1})/p}
    {[\operatorname{Tr}(\mathbf{\Sigma}^{-1})/p]^2}
    - \frac{1}{\operatorname{Tr}(\mathbf{\Xi}^{-1})/p}.
\end{eqnarray}

% Tables~\ref{table1}, \ref{table2}, and \ref{table3} summarize the performance of estimators in terms of the average F and MV loss functions over the $m=1000$, for each of the covariance models, respectively.
% Here, to apply the estimators based on the machine learning approach ResNet, 
% trained using the Adam optimizer with an initial learning rate of $10\times^{-3}$. Internally, the loss function considered to penalize between the predicted and target was the mean squared error (MSE). In addition, a batch of size 16 and 10 epochs was applied to a training data of 100 samples, split with 0.2 for validation. 
Tables~\ref{table1}, \ref{table2}, and \ref{table3} summarize the performance of the estimators in terms of the average $F$ and $MV$ loss functions over the $m = 1000$ Monte Carlo replications, corresponding to each covariance model, respectively. 
For the estimators based on the machine learning approach, the ResNet architecture was trained using the Adam optimizer with an initial learning rate of $10^{-3}$. 
The internal loss function used to penalize the difference between the predicted and target matrices was the mean squared error (MSE). 
Training was performed with a batch size of 16 over 10 epochs, using a training dataset of 100 samples, with 20\% reserved for validation.
\begin{table}[htbp]
\centering
\caption{Block diagonal correlation (model 1). Performance of estimators in terms of $\langle F(\mathbf{\Sigma},\mathbf{\Xi})\rangle$ and $\langle MV(\mathbf{\Sigma},\mathbf{\Xi})\rangle$, where $\langle \cdot\rangle$ represents the average over $m=1000$ realizations of data samples with dimensions $p=100, n=200$.}
\begin{tabular}{|l|r|r|}
\hline
Estimator & \multicolumn{1}{l|}{$\langle F \rangle$} & \multicolumn{1}{l|}{$\langle MV \rangle$} \\ \hline
$\Xi^{naive}$ & 0.507937 & 0.486611 \\ \hline
$\Xi^{LP}$ & 0.065429 & 0.026864 \\ \hline
$\Xi^{CNN}$ & 0.035051 & 0.032022 \\ \hline
$\Xi^{H}$ & 1.054531 & 0.003001 \\ \hline
$\Xi^{ALCA}$ & 0.099357 & 0.057769 \\ \hline
$\Xi^{2S(LP)}$ & 0.056593 & 0.017976 \\ \hline
$\Xi^{2S(CNN)}$ & {\bf 0.034521} & 0.030575 \\ \hline
$\Xi^{2S(H)}$ & 1.054501 & {\bf 0.002551} \\ \hline
\end{tabular}
\label{table1}
\end{table}
\begin{table}[htbp]
\centering
\caption{Completely nested hierarchical covariance (model 2). Performance of estimators in terms of $\langle F(\mathbf{\Sigma},\mathbf{\Xi})\rangle$ and $\langle MV(\mathbf{\Sigma},\mathbf{\Xi})\rangle$, where $\langle \cdot\rangle$ represents the average over $m=1000$ realizations of data samples with dimensions $p=100, n=200$.}
\begin{tabular}{|l|r|r|}
\hline
Estimator & \multicolumn{1}{l|}{$\langle F \rangle$} & \multicolumn{1}{l|}{$\langle MV \rangle$} \\ \hline
$\Xi^{naive}$ & 0.204916 & 0.002551 \\ \hline
$\Xi^{LP}$ & 0.215718 & 0.001520 \\ \hline
$\Xi^{CNN}$ & {\bf 0.186761} & 0.000607 \\ \hline
$\Xi^{H}$ & 3.410518 & {\bf 0.000249} \\ \hline
$\Xi^{ALCA}$ & 0.361030 & 0.009575 \\ \hline
$\Xi^{2S(LP)}$ & 0.370093 & 0.009559 \\ \hline
$\Xi^{2S(CNN)}$ & 0.349912 & 0.007677 \\ \hline
$\Xi^{2S(H)}$ & 3.414401 & 0.000499 \\ \hline
\end{tabular}
\label{table2}
\end{table}
\begin{table}[htbp]
\centering
\caption{Power-law (model 3). Performance of estimators in terms of $\langle F(\mathbf{\Sigma},\mathbf{\Xi})\rangle$ and $\langle MV(\mathbf{\Sigma},\mathbf{\Xi})\rangle$, where $\langle \cdot\rangle$ represents the average over $m=1000$ realizations of data samples with dimensions $p=100, n=200$.}
\begin{tabular}{|l|r|r|}
\hline
Estimator & \multicolumn{1}{l|}{$\langle F \rangle$} & \multicolumn{1}{l|}{$\langle MV \rangle$} \\ \hline
$\Xi^{naive}$ & {\bf 0.000356} & 0.001254 \\ \hline
$\Xi^{LP}$ & 0.000356 & 0.000844 \\ \hline
$\Xi^{CNN}$ & 0.004365 & 0.001281 \\ \hline
$\Xi^{H}$ & 0.006345 & 0.000239 \\ \hline
$\Xi^{ALCA}$ & 0.001992 & 0.010246 \\ \hline
$\Xi^{2S(LP)}$ & 0.001973 & 0.010202 \\ \hline
$\Xi^{2S(CNN)}$ & 0.004852 & 0.002135 \\ \hline
$\Xi^{2S(H)}$ & 0.006345 & {\bf 0.000238} \\ \hline
\end{tabular}
\label{table3}
\end{table}

% For Model 1, the best estimators in terms of $F$ loss is $\mathbf{\Xi}^{2S(CNN)}$, whereas the best performance in terms of $MV$ loss is $\mathbf{\Xi}^{2S(H)}$. 
% In the case of Model 2, the best performance is for $\mathbf{\Xi}^{(CNN)}$ and $\mathbf{\Xi}^{H}$ in terms of $F$ and $MV$ loss, respectively. 
% On the other hand, the performance of Model 3 is particularly intriguing.
% In this case, neither of the estimators really outperforms the naive estimators in terms of the $F$ loss, whereas in terms of the $MV$ loss, most of the estimators struggle to improve the naive one. The only estimators that surpass the naive are $\mathbf{\Xi}^{LP}$, $\mathbf{\Xi}^{H}$, and $\mathbf{\Xi}^{2S(H)}$, being the last the winner overall.
For Model~1, the best-performing estimator in terms of the $F$ loss is $\mathbf{\Xi}^{2S(CNN)}$, whereas the best performance in terms of the $MV$ loss is achieved by $\mathbf{\Xi}^{2S(H)}$. 
In the case of Model~2, the estimators $\mathbf{\Xi}^{CNN}$ and $\mathbf{\Xi}^{H}$ exhibit the lowest $F$ and $MV$ losses, respectively. 
The results for Model~3 are particularly intriguing. 
In this case, none of the estimators significantly outperform the naive estimator in terms of the $F$ loss, while in terms of the $MV$ loss, most estimators struggle to improve upon it. 
The only estimators that surpass the naive benchmark are $\mathbf{\Xi}^{LP}$, $\mathbf{\Xi}^{H}$, and $\mathbf{\Xi}^{2S(H)}$, with the latter showing the overall best performance.

% On the other hand, one can notice that in general both $\mathbf{\Xi}^{H}$ and $\mathbf{\Xi}^{2S(H)}$ do not behave well in terms of $F$ loss but drastically reduce the $MV$ loss. An explanation comes from the fact that the $F$ loss measures all entry errors, and as such, it is sensitive to tiny residual noise. On its part, the $MV$ loss measures effective variance along key directions, and can improve dramatically even if many entries are still slightly noisy. Then, the hybrid estimator is more prone to reducing noise in the key directions given by the eigenvectors of the covariance matrix.
% This phenomenon is more remarkable as the level of structure is more prominent (model 2 and model 3).
On the other hand, it can be observed that, in general, both $\mathbf{\Xi}^{H}$ and $\mathbf{\Xi}^{2S(H)}$ perform poorly in terms of the $F$ loss but substantially reduce the $MV$ loss. 
This behavior can be explained by the fact that the $F$ loss accounts for errors in all individual entries of the covariance matrix, making it highly sensitive to small residual noise. 
In contrast, the $MV$ loss captures deviations along the principal directions of variance, and can therefore exhibit significant improvement even when minor entry-wise noise remains. 
Consequently, the hybrid estimators are more effective at suppressing noise in the dominant eigendirections of the covariance matrix. 
This phenomenon becomes more pronounced as the degree of underlying structure increases, as observed in Models~2 and~3.

\section{Empirical data}
% We consider the daily returns of the main $p=89$ (non-stable) cryptocurrencies in terms of capitalization over 5 years, spanning from 2020-08-02 to 2025-07-31, for a total of $n=1825$ observations. To obtain the sample, we use the API of Yahoo Finance  $\mathtt{yfinance}$. The preprocessing starts with a query of the top 400 cryptocurrencies in terms of capitalization. Then we eliminate the coins that have more than 1\% of missing values and apply the imputation criterion of repeating the last price value to fill the Nans. We also decided to remove the top 10\% cryptos in terms of volatility to have a more reliable dataset. This decision is motivated to avoid the effect of several herding mechanisms like \emph{pump and dump}, or even due to confusion by the euphoric traders choosing the incorrect coin due to a misspell. Finally, we remove all the stablecoins of our remaining dataset (in total 21), like USDT or EURS, because they do not really reflect the dynamics of the cryptocurrency projects, and basically only are designed to maintain a stable value relative to a certain asset, like FIAT money or even commodities like gold. 
We analyze the daily returns of the $p = 89$ major (non-stable) cryptocurrencies by market capitalization over a five-year period, from 2020-08-02 to 2025-07-31, yielding a total of $n = 1825$ observations. The data are retrieved using the \texttt{yfinance} API. The preprocessing pipeline begins by querying the top 400 cryptocurrencies by market capitalization. We then remove all coins with more than 1\% of missing values and impute the remaining gaps using the last observed price (forward fill). To enhance dataset reliability, we further exclude the top 10\% most volatile cryptocurrencies. This filtering step aims to mitigate distortions caused by herding behaviors such as \emph{pump-and-dump} schemes or trader confusion arising from typographical errors in coin identifiers. Finally, we exclude all stablecoins (a total of 21), such as USDT or EURS, as their values are designed to remain pegged to fiat currencies or commodities (e.g., gold) and thus do not reflect the intrinsic dynamics of cryptocurrency markets.

\begin{figure}[hbtp]
    \centering
    \begin{subfigure}{0.45\textwidth}
       \includegraphics[width=\textwidth]{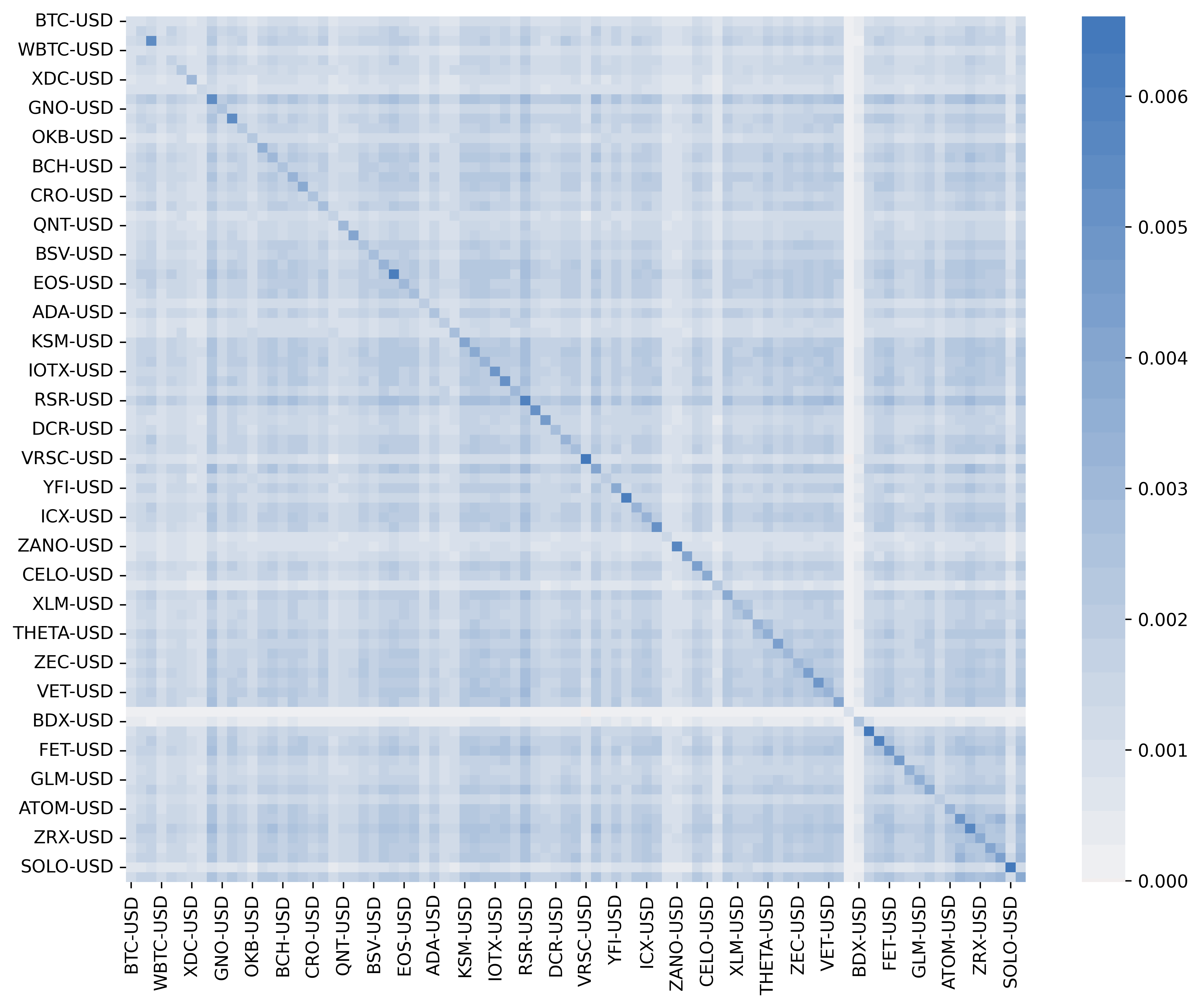}
        \caption{}
    \end{subfigure}\\
    \begin{subfigure}{0.45\textwidth}
       \includegraphics[width=\textwidth]{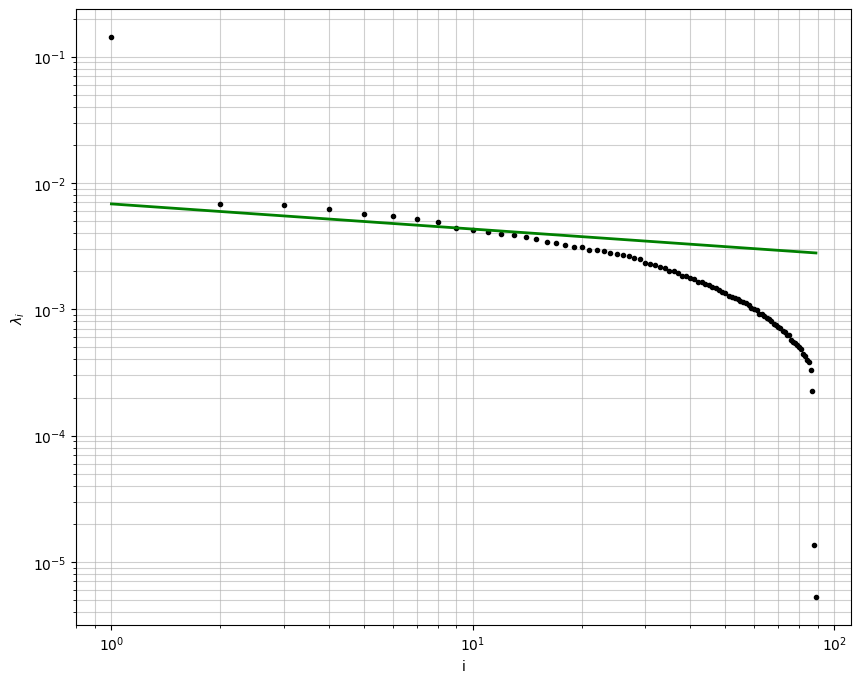}
        \caption{}
    \end{subfigure}\\
    \begin{subfigure}{0.45\textwidth}
        \includegraphics[width=\textwidth]{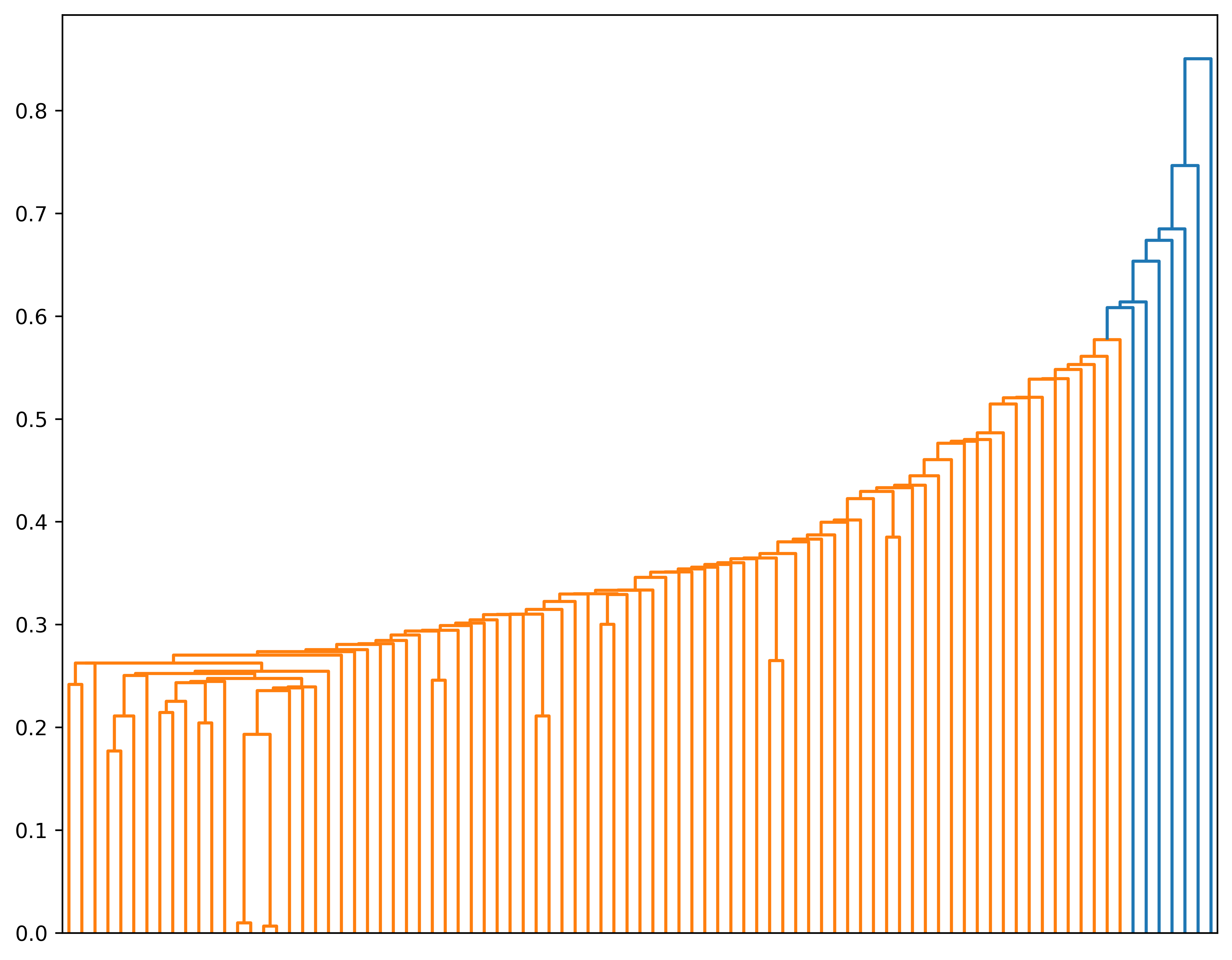}
        \caption{}
    \end{subfigure}
    \caption{Cryptocurrency empirical data (a) Covariance matrix of ordered elements. (b) Scree plot of eigenvalues with a fitted slope of $\alpha=0.2$. (c) Dendrogram under the same methodology as the hierarchical estimator with single linkage (SLCA). For illustrative purpose the descendant links below a cluster node $k$ are equally colored if $k$ is the first node below the cut threshold $t(=0.7)$}
    \label{fig_empirical}
\end{figure}

% In Figure~\ref{fig_empirical}(a), the covariance matrix of the cryptocurrency empirical dataset is shown. Here we have ordered the elements applying the seriation algorithm of~\cite{atkins1998spectral} for a twofold purpose: (1) better graphical representation, where similar covariances are close to each other, (2) improve capabilities of the deep learning estimators to capture and learn the structure of the noise covariance matrices. We can notice that the structure of the ordered covariance matrix presents similar patterns as those proposed by the powerlaw model.
In Figure~\ref{fig_empirical}(a), we present the empirical covariance matrix of the cryptocurrency dataset. The assets have been ordered using the seriation algorithm proposed by~\cite{atkins1998spectral} with a twofold objective: (i) to enhance the graphical representation by positioning assets with similar covariance patterns closer to each other, and (ii) to improve the ability of deep learning estimators to detect and learn the structural dependencies within the noisy covariance matrices. The resulting ordered covariance matrix exhibits patterns that closely resemble those generated by the power-law model, suggesting that the empirical structure of the cryptocurrency market may share similar scaling properties.
% Figure~\ref{fig_empirical}(b) shows the screeplot of eigenvalues of the covariance matrix. We can see that the behavior resembles that of models 2 and 3, except for some outliers in the tails. Also, the dendrogram in Figure~\ref{fig_empirical}~(c) shows a pattern that qualitatively matches the one of models 2 and 3. Then, our covariance models capture these stylized facts observed in the empirical dataset of cryptocurrencies.
Figure~\ref{fig_empirical}(b) displays the scree plot of the eigenvalues of the empirical covariance matrix. The observed behavior closely resembles that of Models~2 and~3, with the exception of a few outliers in the tails. The fitted slope of $\alpha=0.2$ is superimposed as a graphical reference. Moreover, the dendrogram presented in Figure~\ref{fig_empirical}(c) exhibits a hierarchical pattern qualitatively consistent with the structures observed in these two models. These results indicate that our proposed covariance models effectively capture the stylized facts present in the empirical cryptocurrency dataset.

% Then we split our data into a train and a test period. The splitting point is chosen to be 2021-11-09 because it reached the maximum closing value of BTC during the pandemic turmoil\footnote{The peak was exactly on November 10, 2021, but we are considering the midnight value as the closing price.} and after that the prices start to decline to minimum local. The bear market bottomed in Nov 2022 (~$15.5k during the FTX crisis).In this way, we are putting a limit to learn in a bull market and trying to apply it to bear market conditions.
We split our dataset into training and testing periods. The splitting point is set to 2021-11-09, corresponding to the maximum closing value of Bitcoin (BTC) during the pandemic turmoil.\footnote{The absolute peak occurred on November 10, 2021; however, we consider the midnight value as the closing price.} Following this date, prices began to decline, reaching a local minimum in November 2022 during the FTX crisis. This partition enables the model to learn under bull market conditions and to be subsequently evaluated under bear market dynamics, thereby facilitating an assessment of its robustness across contrasting market regimes.

% Then, starting on 2021-11-09, a walk-forward analysis (rolling window backtesting) was performed using an in-sample and out-of-sample covariance matrix of $N=182$ trading days and rebalanced also every $N$ days. In total, we rebalance the portfolio seven times to cover an investment period from 2021-11-09 to 2025-05-05. 
Starting from 2021-11-09, we conducted a walk-forward analysis (rolling-window backtesting) using in-sample and out-of-sample covariance matrices computed over $N=182$ trading days, with portfolio rebalancing performed at the same frequency. In total, the portfolio was rebalanced seven times, covering the investment period from 2021-11-09 to 2025-05-05.
% Thus, we obtain the optimal portfolio allocations $\mathbf{W}$, considering as an input of the MV+ investment strategy each of the covariance estimators presented in section~\ref{estimators} applied to the in-sample data. 
% In the case of the ResNet estimators ($\mathbf{\Xi}^{CNN}$, $\mathbf{\Xi}^{2S(CNN)}$, $\mathbf{\Xi}^{H}$, and $\mathbf{\Xi}^{2S(H)}$) we train the model each rebalancing time using an extended training dataset that cover the in-sample data plus almost an additional past year (282 days). The idea was to start training using the first observation of our complete dataset and apply the same moving window criterion to always have the same size of training points in each rebalancing scenario.
Thus, we obtain the optimal portfolio allocations $\mathbf{w}$ by applying the MVP+ investment strategy, using as input each of the covariance estimators presented in Section~\ref{estimators}, computed over the in-sample data.
For the ResNet-based estimators ($\mathbf{\Xi}^{CNN}$, $\mathbf{\Xi}^{2S(CNN)}$, $\mathbf{\Xi}^{H}$, and $\mathbf{\Xi}^{2S(H)}$), the model is retrained at each rebalancing point using an extended training dataset that includes the in-sample period plus approximately one additional preceding year (282 days). The rationale is to initiate training with the first observation of the complete dataset and subsequently apply a consistent rolling-window scheme, maintaining a constant number of training observations across all rebalancing scenarios.

\begin{figure}[hbtp]
    \centering
    \begin{subfigure}{0.85\textwidth}
        \includegraphics[width=\textwidth]{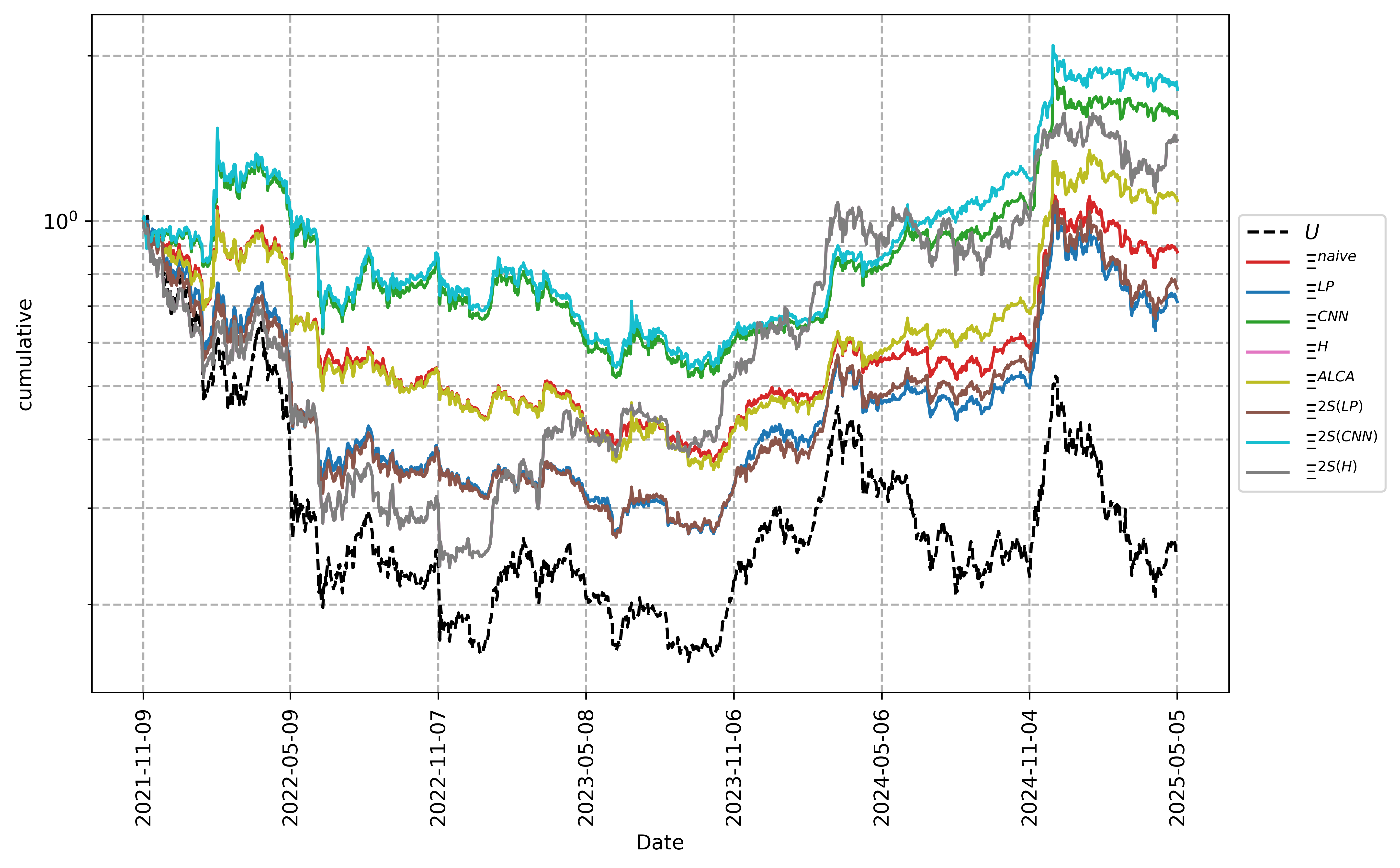}
    \caption{}
    \end{subfigure}\\
    \begin{subfigure}{0.85\textwidth}
        \includegraphics[width=\textwidth]{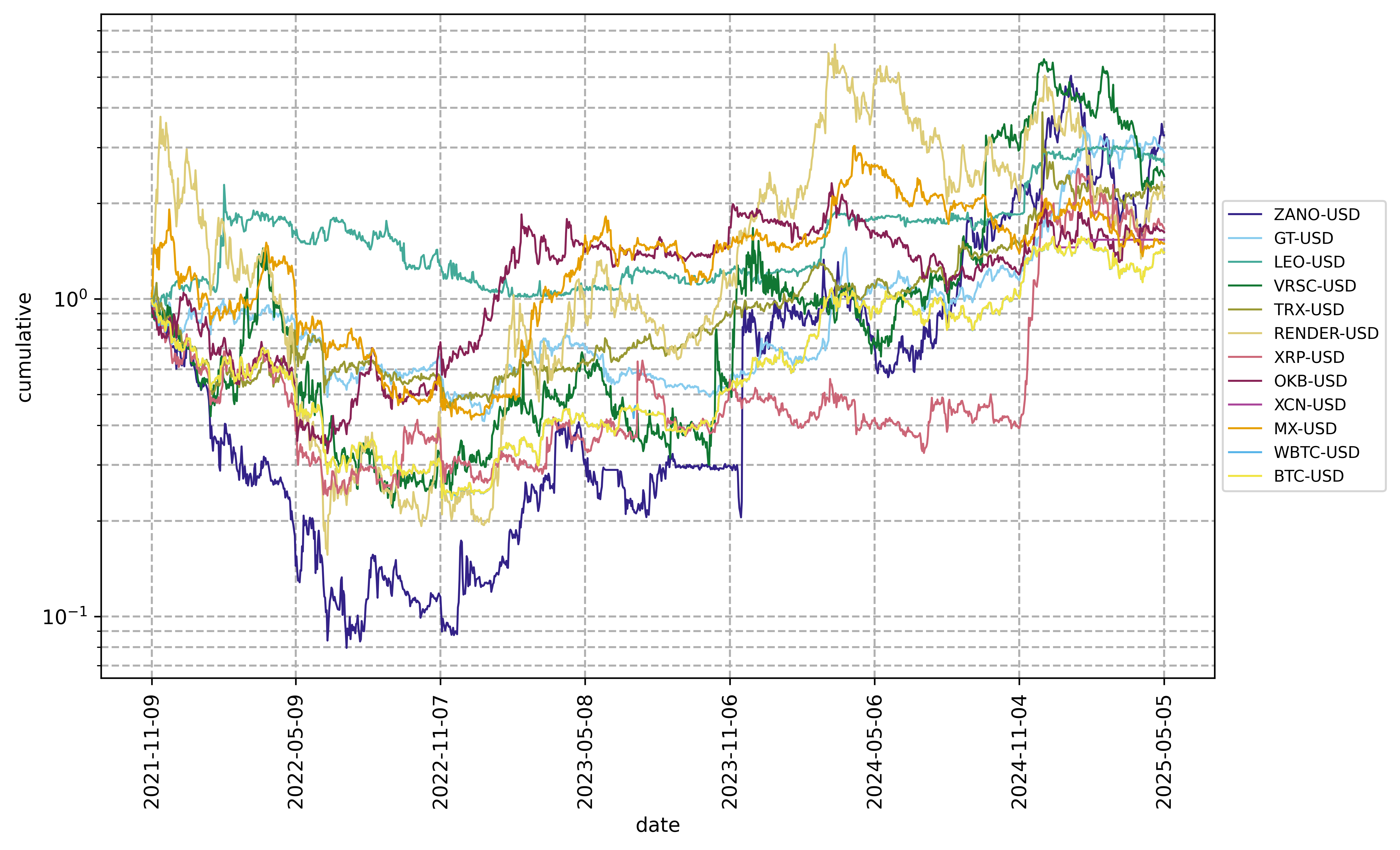}
    \caption{}
    \end{subfigure}
    \caption{(a) Walk-forward cumulative returns on empirical data for MVP+. The weights are optimized with $T_{in}=182$ days, applied over $T_{out}=182$, and rebalancing every $\Delta T = 182$ days. (b) Cumulative returns of the top 12-performing individual cryptocurrencies under the buy-and-hold strategy. The vertical axis is on a logarithmic scale for better visualization.}
    \label{fig_walkforward}
\end{figure}

% In this way, using a stride of 1, we obtain a training sample of 100 datasets of dimension $p\times N$ to apply a learning process. Here, the setting for the ResNet algorithm is the same as that used in the simulation. 
% Next, we compute the portfolio returns by multiplying $\mathbf{W}_{in}$ with the out-of-sample returns $\mathbf{R}_{out}$.
% Figure~\ref{fig_walkforward}~(a)  shows the cumulative return for each estimator during the entire walk-forward analysis; the rebalancing dates have been delimited by grey dashed vertical lines.
Using a stride of 1, we thus obtain a training sample of 100 datasets, each of dimension $p \times N$, to apply the learning process. The ResNet algorithm is configured using the same settings as in the simulation study. Next, out-of-sample portfolio returns~( $\mathbf{R}_{\text{out}}$) are computed by multiplying the in-sample allocations  ($\mathbf{w}_{\text{in}}^{\top}$) with the out-of-sample returns ($\mathbf{r}_{\text{out}}$):
\begin{equation}
    \mathbf{R}_{t}^{\text{out}} = \mathbf{w}_{\text{in}}^{\top} \mathbf{r}_{\text{out}} .  
\end{equation} 
Figure~\ref{fig_walkforward}(a) displays the cumulative returns for each estimator throughout the walk-forward analysis, with the rebalancing dates indicated by grey dashed vertical lines.
%We can also see in figure~\ref{fig_walkforward}~(b) the top-12 performance individual cryptocurrencies during the same analyzed period. It can be seen that the performance of individual cryptos behaves more erratically (risky) and consequently can reach a higher cumulative return.
Figure~\ref{fig_walkforward}(b) shows the performance of the top 12 individual cryptocurrencies over the same period. The returns of individual cryptocurrencies exhibit greater volatility, reflecting higher risk, and consequently they can achieve higher cumulative returns compared to the diversified portfolios.

% Table~\ref{table4} shows the walk-forward portfolio performance in terms of financial metrics for each of the covariance estimators. As a reference strategy it has been included the performance of a uniform portfolio ($U$), i.e., an investment portfolio where the capital is equally distributed in all the assets.
Table~\ref{table4} presents the walk-forward portfolio performance across various financial metrics for each of the covariance estimators. As a benchmark, we also include the performance of a uniform portfolio ($U$), in which capital is equally allocated across all assets.
\begin{table}[hbtp]
\footnotesize
\centering
\begin{tabular}{|l|l|l|l|l|l|l|}
\hline
Estimator & \parbox[t]{1.6cm}{Cumulative\\ Return\\} & \parbox[t]{1.2cm}{Annual \\ Return} & \parbox[t]{1.5cm}{Annual \\ Volatility} & \parbox[t]{1cm}{SR} & \parbox[t]{1.5cm}{Maximum \\ Drawdown} & Turnover \\
\hline
$U$ & 0.25 & -33.00\% & 66.52\% & -0.5 & -84.56\% & 0 \\ \hline
$\Xi^{naive}$ & 0.88 & -3.66\% & {\bf 39.10\%} & -0.09 & -65.63\% & 1.02 \\ \hline
$\Xi^{LP}$ & 0.71 & -9.24\% & 43.75\% & -0.21 & -73.35\% & 1.13 \\ \hline
$\Xi^{CNN}$ & 1.54 & 13.19\% & 46.79\% & 0.28 & -63.74\% & 1.3 \\ \hline
$\Xi^{H}$ & 1.4 & 10.17\% & 54.44\% & 0.19 & -76.43\% & 0 \\ \hline
$\Xi^{ALCA}$ & 1.09 & 2.46\% & 40.71\% & 0.06 & -66.24\% & 1 \\ \hline
$\Xi^{2S(LP)}$ & 0.75 & -7.78\% & 43.99\% & -0.18 & -73.75\% & 1.22 \\ \hline
$\Xi^{2S(CNN)}$ & {\bf 1.74} & {\bf 17.14\%} & 46.60\% & {\bf 0.37} & {\bf -63.63\%} & 1.34 \\ \hline
$\Xi^{2S(H)}$ & 1.4 & 10.17\% & 54.44\% & 0.19 & -76.43\% & 0 \\ \hline
\end{tabular}
\caption{Walk-forward portfolio performance for the cryptocurrency market under the MVP+ investment strategy. The weights are optimized with $T_{in}=182$ days, applied over $T_{out}=182$, and rebalancing every $\Delta T = 182$ days.}
\label{table4}
\end{table}
% It can be noticed that the highest cumulative portfolio return is under the $\Xi^{2S(CNN)}$ estimator. By analyzing the number of annual returns, which are computed as the geometric mean of the portfolio return multiplied by the factor 365 to obtain a proxy of yearly quantities, it is notice that only the \emph{ResNet} estimators and \emph{ALCA}  achieve profits; the other obtain cumulative values less than 1, i.e, the positive values correspond to the \emph{ResNet} and \emph{ALCA} estimators, while the rest are negative. 
It can be observed that the highest cumulative portfolio return is achieved using the $\mathbf{\Xi}^{2S(CNN)}$ estimator. By examining annualized returns, computed as the geometric mean of portfolio returns scaled by a factor of 365 to approximate yearly performance, we find that only the \emph{ResNet}-based estimators and the \emph{ALCA} estimator generate positive returns (see third column).

The annual portfolio volatility is measured by eq.~\ref{volatility} and also multiplied by the factor 365 to obtain a proxy. Interestingly, the \emph{naive} estimator reaches the minimum value under this metric. Thus, an increase in mean return does not necessarily imply a decrease in standard deviation.  The Sharpe Ratio~(SR), defined as the ratio between the mean and standard deviation of the portfolio return, gives us a fairer number to compare different investments. Under this measure, the most equilibrated portfolio is obtained again by the $\mathbf{\Xi}^{2S(CNN)}$ estimator.

The maximum drawdown measures the largest peak-to-trough decline in the cumulative return of the investment portfolio expressed as a percentage. This measure represents the maximum loss presented in the investment.
Also, the $\mathbf{\Xi}^{2S(CNN)}$ improve the loss to $-63.63\%$. Yet, the improvement is marginal in relation to the \emph{naive} estimator, which is of $-65.65\%$. Finally, turnover measures the absolute difference between the asset weights at consecutive periods, normalized by the number of periods. A high value is associated with high transaction fees. This number ranges between 0 and 2, with 1 meaning that half of the assets reallocate in every rebalancing. In this case, the $\mathbf{\Xi}^{H}$ and $\mathbf{\Xi}^{2S(H)}$ estimators do not involve any reallocation. In fact, this particular strategy allocates 100\% of the portfolio to BTC-USD at all times.

Figure~\ref{table5} shows the equivalent financial metrics but for the buy \& hold strategy of investing in individual cryptocurrencies through the walk-forward period. Notice that here, the turnover metric is not measured because we never rebalance. Interestingly, several coins surpass the performance of the combination of a state-of-the-art covariance estimator and the classical investment strategy MVP+. Even accounting for the trade-off measured by the SR and the Maximum Drawdown, crypto like GT-USD outperforms all strategies. 
Therefore, an important direction for future research is to integrate asset selection into the overall covariance estimation and portfolio allocation framework.
\begin{table}[hbtp]
\footnotesize
\centering
\begin{tabular}{|l|l|l|l|l|l|l|}
\hline
Cryptocurrency & \parbox[t]{1.8cm}{Cumulative\\ Return\\} & \parbox[t]{1.2cm}{Annual \\ Return} & \parbox[t]{1.5cm}{Annual \\ Volatility} & \parbox[t]{1.3cm}{SR} & \parbox[t]{1.5cm}{Maximum \\ Drawdown}\\
\hline
ZANO-USD & {\bf 3.28} & {\bf 40.50\%} & 134.62\% & 0.30 & -92.06\% \\ \hline
GT-USD & 2.90 & 35.61\% & 53.50\% & {\bf 0.67} & -62.41\% \\ \hline
LEO-USD & 2.64 & 32.13\% & 52.74\% & 0.61 & {\bf -55.67\%} \\ \hline
VRSC-USD & 2.44 & 29.09\% & 132.76\% & 0.22 & -84.75\% \\ \hline
TRX-USD & 2.26 & 26.28\% & 65.97\% & 0.40 & -59.52\%  \\ \hline
RENDER-USD & 2.11 & 23.87\% & 129.75\% & 0.18 & -95.83\%  \\ \hline
XRP-USD & 1.67 & 15.73\% & 84.43\% & 0.19 & -75.53\%  \\ \hline
OKB-USD & 1.62 & 14.91\% & 67.45\% & 0.22 & -68.81\%  \\ \hline
XCN-USD & 1.54 & 13.10\% & {\bf 52.05\%} & 0.25 & -76.42\%  \\ \hline
MX-USD & 1.49 & 12.17\% & 66.73\% & 0.18 & -78.22\%  \\ \hline
WBTC-USD & 1.40 & 10.19\% & 54.27\% & 0.19 & -76.55\%  \\ \hline
BTC-USD & 1.40 & 10.17\% & 54.44\% & 0.19 & -76.43\% \\ \hline
\end{tabular}
\caption{Walk-forward portfolio performance for the top 12 individual cryptocurrencies under the buy \& hold strategy.}
\label{table5}
\end{table}

\section{Conclusion}
% Hemos encontrado que el estimador a dos pasos basado en la arquitectura ResNet $(\mathbf{\Xi^{2S(CNN)}})$ minimiza la funcion de perdida de Frobenius para el modelo de bloques diagonales. Para este mismo modelo el estimador de dos pasos hibrido  $(\mathbf{\Xi^{2S(H)}})$ minimiza la funcion de perdida de minima varianza. Cuando incrementamos la complejidad del modelo de covarianza encontramos que no es necesario aplicar el segundo filtro, oor lo que para estas  mismas funciones de perdida los estimadores  $(\mathbf{\Xi^{CNN}})$ y  $(\mathbf{\Xi^{H}})$ son suficientes para reducir el ruido al minimo, de manera respectiva.
% Por otro lado, un fenomeno interesante surge cuando el modelo de covarianza es del tipo de ley de potencia, en este caso ningun estimador del estado del arte mejora la estimacion naive en relacion a Frobenius. Pero en el caso de la funcion de minima varianza el estimador de dos pasos hibrido lleva al minim el ruido. El patron general que observamos es que la funcion de perdida de minima varianza se reduce de manera  sistematica en todos los modelos al considerar los estimadores basados en ResNet. 
% Here the distintive point is that MV loss captures deviations along the principal directions of variance, and as as such are more adequate to portfolio applications. 

We found that the two-step estimator based on the ResNet architecture ($\mathbf{\Xi}^{2S(CNN)}$) minimizes the Frobenius loss function for the block-diagonal model. For the same model, the two-step hybrid estimator ($\mathbf{\Xi}^{2S(H)}$) attains the minimum value of the MV loss function. As the complexity of the covariance model increases, we observe that applying the second filtering step becomes unnecessary; in such cases, the single-step estimators ($\mathbf{\Xi}^{CNN}$) and ($\mathbf{\Xi}^{H}$) are sufficient to minimize noise with respect to the Frobenius and MV losses, respectively. 
Moreover, an interesting phenomenon arises when the covariance model follows a power-law structure: in this case, none of the state-of-the-art estimators outperform the naive one in terms of the Frobenius loss. However, regarding the MV loss, the two-step hybrid estimator achieves the lowest noise level. Overall, a consistent pattern emerges MV loss systematically decreases across all models when using the ResNet-based estimators.
The distinctive point here is that the MV loss captures deviations along the principal directions of variance and, as such, is more suitable for portfolio optimization applications.
In sum, as market structure becomes more complex (hierarchical or power-law), hybrid estimators increasingly outperform traditional methods in terms of portfolio risk control.

On the other hand, when considering empirical data from the cryptocurrency market, we observe that the covariance structure, the associated scree plot, and the dendrogram, exhibit stylized facts closely resembling those found in both the nested hierarchical covariance model and the power-law covariance model. In particular, the eigenvalue spectrum displays an approximate power-law decay with coefficient $\alpha = 0.2$. These observations provide initial confirmation of the conjecture proposed in~\cite{garcia2025high}, namely that the nested hierarchical covariance model adequately characterizes the complex dynamics of financial instruments—here extended to the cryptocurrency market. Moreover, the empirical evidence supports a natural generalization of this conjecture to a broader class of covariance structures governed by power-law behavior.

Focusing on the specific dataset analyzed, we find that the best financial performance metrics are generally obtained using the two-step CNN-based estimator $\boldsymbol{\Xi}^{2S(\mathrm{CNN})}$ within the MVP+ portfolio framework. Notably, this result remains robust under the deliberately challenging experimental design, in which the training sample corresponds to a bull-market regime, while the out-of-sample evaluation spans a heterogeneous period beginning with a pronounced bear market. Empirical evidence further indicates that cryptocurrency returns are dominated by a strong market-wide factor, which is preserved by all covariance estimators and leads to persistently high out-of-sample volatility. As a consequence, covariance estimation primarily affects the orientation of risk—through modifications of eigenvectors and relative factor loadings—rather than the overall volatility level. Within this context, CNN-based estimators, and in particular the hybrid $\boldsymbol{\Xi}^{2S(\mathrm{CNN})}$, consistently deliver higher cumulative returns, superior Sharpe ratios, and reduced drawdowns, despite exhibiting volatility levels comparable to those of competing approaches. By contrast, the hybrid estimator $\boldsymbol{\Xi}^{\mathrm{H}}$ largely mirrors the behavior of Bitcoin itself, reflecting its strong alignment with the dominant market mode. While RMT-based and ALCA estimators improve portfolio stability or reduce turnover, they fail to match the return and drawdown performance achieved by the CNN-based methods.

Nevertheless, although these results are promising in terms of adequately characterizing the complex dynamics of cryptocurrencies and in proposing a hybrid estimator that combines state-of-the-art neural network architectures with the analytical foundations of random matrix theory, they are not sufficient to achieve the maximum cumulative return. When compared with the performance of individual assets, certain instruments are capable of nearly tripling the initial capital while exhibiting a more favorable balance between risk and return, as measured by the Sharpe ratio. This indicates that, from a practical perspective, further improvements remain possible. In particular, future performance gains are likely to require the continued integration of complementary methodologies and the inclusion of an ingredient that remains relatively uncommon within econophysics—namely, asset selection and valuation grounded in fundamental analysis~\cite{graham2003intelligent}. This challenge is especially pronounced in cryptocurrency markets, where traditional fundamental metrics are largely absent.

From a statistical-physics perspective, the proposed hybrid framework extends well beyond financial applications and naturally fits within the broader class of inference problems involving low-rank deformations of Wishart ensembles. In such systems, information is not solely contained in spectral outliers but also in the collective organization and alignment of eigenvectors, which undergo sharp detectability thresholds governed by the Baik–Ben Arous–Péché (BBP) phase transition~\cite{baik2005phase, bloemendal2013limits}. Below the BBP threshold, informative factors are statistically indistinguishable from noise at the level of eigenvalues, while partial but nontrivial information persists in the eigenvector overlaps. By explicitly exploiting this eigenvector structure, the hybrid estimator effectively probes the vicinity of the BBP transition, enabling the detection of multiscale and hierarchical organization that remains inaccessible to purely eigenvalue-based denoising methods.
\medskip

\noindent{\bf Competing Interests:} The authors have no competing interests to declare that are relevant to the content of this article.
\medskip

\noindent{\bf Data availability statement:} The data that support the findings of this study are available from the corresponding author upon request.
\medskip

\noindent
{\bf Declaration of generative AI and AI-assisted technologies in the manuscript preparation process.}
During the preparation of this work, the author used ChatGPT to assist with readability improvements and proofreading. After using this tool/service, the author reviewed and edited the content as needed and takes full responsibility for the content of the published article.
\medskip

\noindent
{\bf Author Contribution Statement.}
The author A.G.M. confirms being the sole contributor to this work and has approved it for publication.
The author was responsible for the conception and design of the study, data collection, analysis, interpretation of results, and the writing and revision of the manuscript.

%%%%%%%%%%%%%%%%%%%%%%%%%%%%%%%%%%%%%%%%%%%%%%%%%%%%%%%%%%%%%%%%%%%%%%%%%%%%%%%%%%%%%%%%
\printbibliography

@article{tumminello2007hierarchically,
  title={Hierarchically nested factor model from multivariate data},
  author={Tumminello, Michele and Lillo, Fabrizio and Mantegna, Rosario N},
  journal={EPL (Europhysics Letters)},
  volume={78},
  number={3},
  pages={30006},
  year={2007},
  publisher={IOP Publishing}
}

@article{ledoit2022power,
  title={The power of (non-) linear shrinking: A review and guide to covariance matrix estimation},
  author={Ledoit, Olivier and Wolf, Michael},
  journal={Journal of Financial Econometrics},
  volume={20},
  number={1},
  pages={187--218},
  year={2022},
  publisher={Oxford University Press}
}

@article{tumminello2007kullback,
  title={Kullback-Leibler distance as a measure of the information filtered from multivariate data},
  author={Tumminello, Michele and Lillo, Fabrizio and Mantegna, Rosario N},
  journal={Physical Review E},
  volume={76},
  number={3},
  pages={031123},
  year={2007},
  publisher={APS}
}

@article{laloux1999noise,
  title={Noise dressing of financial correlation matrices},
  author={Laloux, Laurent and Cizeau, Pierre and Bouchaud, Jean-Philippe and Potters, Marc},
  journal={Physical review letters},
  volume={83},
  number={7},
  pages={1467},
  year={1999},
  publisher={APS}
}

@article{bun2017cleaning,
  title={Cleaning large correlation matrices: tools from random matrix theory},
  author={Bun, Jo{\"e}l and Bouchaud, Jean-Philippe and Potters, Marc},
  journal={Physics Reports},
  volume={666},
  pages={1--109},
  year={2017},
  publisher={Elsevier}
}

@article{johnstone2001distribution,
  title={On the distribution of the largest eigenvalue in principal components analysis},
  author={Johnstone, Iain M},
  journal={The Annals of statistics},
  volume={29},
  number={2},
  pages={295--327},
  year={2001},
  publisher={Institute of Mathematical Statistics}
}

@article{ledoit2011eigenvectors,
  title={Eigenvectors of some large sample covariance matrix ensembles},
  author={Ledoit, Olivier and P{\'e}ch{\'e}, Sandrine},
  journal={Probability Theory and Related Fields},
  volume={151},
  number={1},
  pages={233--264},
  year={2011},
  publisher={Springer}
}

@article{ledoit2020analytical,
  title={Analytical nonlinear shrinkage of large-dimensional covariance matrices},
  author={Ledoit, Olivier and Wolf, Michael},
  journal={The Annals of Statistics},
  volume={48},
  number={5},
  pages={3043--3065},
  year={2020},
  publisher={Institute of Mathematical Statistics}
}

@article{Burda2022,
  title={Cleaning large-dimensional covariance matrices for correlated samples},
  author={Burda, Zdzislaw and Jarosz, Andrzej},
  journal={Physical Review E},
  volume={105},
  number={3},
  pages={034136},
  year={2022},
  publisher={APS}
}

@article{ledoit2004well,
  title={A well-conditioned estimator for large-dimensional covariance matrices},
  author={Ledoit, Olivier and Wolf, Michael},
  journal={Journal of multivariate analysis},
  volume={88},
  number={2},
  pages={365--411},
  year={2004},
  publisher={Elsevier}
}

@book{potters2020first,
  title={A First Course in Random Matrix Theory: For Physicists, Engineers and Data Scientists},
  author={Potters, Marc and Bouchaud, Jean-Philippe},
  year={2020},
  publisher={Cambridge University Press}
}

@book{johnson2002applied,
  title={Applied multivariate statistical analysis},
  author={Johnson, Richard Arnold and Wichern, Dean W},
  year={2002},
  publisher={Pearson Prentice Hall}
}

@article{donoho2018optimal,
  title={Optimal shrinkage of eigenvalues in the spiked covariance model},
  author={Donoho, David L and Gavish, Matan and Johnstone, Iain M},
  journal={Annals of statistics},
  volume={46},
  number={4},
  pages={1742},
  year={2018},
  publisher={NIH Public Access}
}

@article{tola2008cluster,
  title={Cluster analysis for portfolio optimization},
  author={Tola, Vincenzo and Lillo, Fabrizio and Gallegati, Mauro and Mantegna, Rosario N},
  journal={Journal of Economic Dynamics and Control},
  volume={32},
  number={1},
  pages={235--258},
  year={2008},
  publisher={Elsevier}
}

@article{bongiorno2021covariance,
  title={Covariance matrix filtering with bootstrapped hierarchies},
  author={Bongiorno, Christian and Challet, Damien},
  journal={PloS one},
  volume={16},
  number={1},
  pages={e0245092},
  year={2021},
  publisher={Public Library of Science San Francisco, CA USA}
}

@article{bongiorno2022reactive,
  title={Reactive global minimum variance portfolios with k-BAHC covariance cleaning},
  author={Bongiorno, Christian and Challet, Damien},
  journal={The European Journal of Finance},
  volume={28},
  number={13-15},
  pages={1344--1360},
  year={2022},
  publisher={Taylor \& Francis}
}

@article{pantaleo2011improved,
  title={When do improved covariance matrix estimators enhance portfolio optimization? An empirical comparative study of nine estimators},
  author={Pantaleo, Ester and Tumminello, Michele and Lillo, Fabrizio and Mantegna, Rosario N},
  journal={Quantitative Finance},
  volume={11},
  number={7},
  pages={1067--1080},
  year={2011},
  publisher={Taylor \& Francis}
}

@inproceedings{tumminello2007spectral,
  title        = {Spectral properties of correlation matrices for hierarchically nested factor models},
  author       = {Tumminello, Michele and Lillo, Fabrizio and Mantegna, Rosario N.},
  booktitle    = {AIP Conference Proceedings},
  volume       = {965},
  number       = {1},
  pages        = {300--307},
  year         = {2007},
  publisher    = {American Institute of Physics},
  address      = {Melville, NY}
}

@article{garcia2023two,
  title={Two-step estimators of high-dimensional correlation matrices},
  author={Garc{\'\i}a-Medina, Andr{\'e}s and Miccich{\`e}, Salvatore and Mantegna, Rosario N},
  journal={Physical Review E},
  volume={108},
  number={4},
  pages={044137},
  year={2023},
  publisher={APS}
}

@book{roncalli2013introduction,
  title={Introduction to risk parity and budgeting},
  author={Roncalli, Thierry},
  year={2013},
  publisher={CRC Press}
}

@article{yang2015robust,
  title={A robust statistics approach to minimum variance portfolio optimization},
  author={Yang, Liusha and Couillet, Romain and McKay, Matthew R},
  journal={IEEE Transactions on Signal Processing},
  volume={63},
  number={24},
  pages={6684--6697},
  year={2015},
  publisher={IEEE}
}

@book{boyd2004convex,
  title={Convex optimization},
  author={Boyd, Stephen and Vandenberghe, Lieven},
  year={2004},
  publisher={Cambridge university press}
}

@article{garcia2024random,
  title={Random Matrix Theory and Nested Clustered Optimization on high-dimensional portfolios},
  author={Garc\'ia-Medina, Andrés and Rodr\'iguez-Camejo, Benito},
  journal={International Journal of Modern Physics C},
  volume={35},
  number={08},
  pages={1--19},
  year={2024},
  publisher={World Scientific Publishing Co. Pte. Ltd.}
}

@article{markowitz1952,
  title={Portfolio selection},
  author={Harry Markowitz},
  journal={The journal of finance},
  volume={7},
  number={1},
  pages={77--91},
  year={1952},
  publisher={Wiley}
}

@article{cahill2004fibonacci,
  title={Fibonacci and Lucas numbers as tridiagonal matrix determinants},
  author={Cahill, Nathan D and Narayan, Darren A},
  journal={The Fibonacci Quarterly},
  volume={42},
  number={3},
  pages={216--221},
  year={2004},
  publisher={Taylor \& Francis}
}

@article{garcia2025high,
  title={High-dimensional covariance matrix estimators on simulated portfolios with complex structures},
  author={Garc{\'\i}a-Medina, Andr{\'e}s},
  journal={Physical Review E},
  volume={111},
  number={2},
  pages={024316},
  year={2025},
  publisher={APS}
}

@article{bongiorno2025end,
  title        = {End-to-End Large Portfolio Optimization for Variance Minimization with Neural Networks through Covariance Cleaning},
  author       = {Bongiorno, Christian and Manolakis, Efstratios and Mantegna, Rosario Nunzio},
  journal      = {arXiv preprint arXiv:2507.01918},
  year         = {2025},
  month        = jul,
  note         = {Submitted 2 Jul 2025},
  eprint       = {2507.01918},
  primaryClass = {q-fin.PM},
}

@article{atkins1998spectral,
  title={A spectral algorithm for seriation and the consecutive ones problem},
  author={Atkins, Jonathan E and Boman, Erik G and Hendrickson, Bruce},
  journal={SIAM Journal on Computing},
  volume={28},
  number={1},
  pages={297--310},
  year={1998},
  publisher={SIAM}
}

@inproceedings{he2016deep,
  title={Deep residual learning for image recognition},
  author={He, Kaiming and Zhang, Xiangyu and Ren, Shaoqing and Sun, Jian},
  booktitle={Proceedings of the IEEE conference on computer vision and pattern recognition},
  pages={770--778},
  year={2016}
}

@article{Watorek2021Multiscale,
  title        = {Multiscale characteristics of the emerging global cryptocurrency market},
  author       = {W\k{a}torek, Marcin and Dro\.z{}d\.z{}, Stanis\l{}aw and Kwapie\'n, Jaros\l{}aw and Minati, Ludovico and O\'swi\k{e}cimka, Pawe\l{} and Stanuszek, Marek},
  journal      = {Physics Reports},
  volume       = {901},
  pages        = {1--82},
  year         = {2021},
  publisher    = {Elsevier},
  doi          = {10.1016/j.physrep.2020.10.005}
}

@book{mehta2004random,
  title={Random matrices},
  author={Mehta, Madan Lal},
  volume={142},
  year={2004},
  publisher={Elsevier}
}

@article{bohigas1984characterization,
  title={Characterization of chaotic quantum spectra and universality of level fluctuation laws},
  author={Bohigas, Oriol and Giannoni, Marie-Joya and Schmit, Charles},
  journal={Physical review letters},
  volume={52},
  number={1},
  pages={1},
  year={1984},
  publisher={APS}
}

@article{beenakker1997random,
  title={Random-matrix theory of quantum transport},
  author={Beenakker, Carlo WJ},
  journal={Reviews of modern physics},
  volume={69},
  number={3},
  pages={731},
  year={1997},
  publisher={APS}
}

@article{potters2005financial,
  title={Financial applications of random matrix theory: Old laces and new pieces},
  author={Potters, Marc and Bouchaud, Jean-Philippe and Laloux, Laurent},
  journal={arXiv preprint physics/0507111},
  year={2005}
}

@article{he2020resnet,
  title={Why resnet works? residuals generalize},
  author={He, Fengxiang and Liu, Tongliang and Tao, Dacheng},
  journal={IEEE transactions on neural networks and learning systems},
  volume={31},
  number={12},
  pages={5349--5362},
  year={2020},
  publisher={IEEE}
}

@article{makridakis2024avoiding,
  title     = {Avoiding overconfidence: Evidence from the M6 financial competition},
  author    = {Makridakis, Spyros and Spiliotis, Evangelos and Michailidis, Maria},
  journal   = {International Journal of Forecasting},
  volume    = {41},
  number    = {4},
  pages     = {1395--1403},
  year      = {2025},
  publisher = {Elsevier},
  doi       = {10.1016/j.ijforecast.2024.10.001}
}

@article{mattera2023shrinkage,
  title={Shrinkage estimation with reinforcement learning of large variance matrices for portfolio selection},
  author={Mattera, Giulio and Mattera, Raffaele},
  journal={Intelligent Systems with Applications},
  volume={17},
  pages={200181},
  year={2023},
  publisher={Elsevier}
}

@book{brownlee2018deep,
  title={Deep learning for time series forecasting: predict the future with MLPs, CNNs and LSTMs in Python},
  author={Brownlee, Jason},
  year={2018},
  publisher={Machine Learning Mastery}
}

@book{nielsen2015neural,
  title={Neural networks and deep learning},
  author={Nielsen, Michael A},
  volume={25},
  year={2015},
  publisher={Determination press San Francisco, CA, USA}
}

@book{graham2003intelligent,
  title={The intelligent investor},
  author={Graham, Benjamin and Zweig, Jason},
  year={2003},
  publisher={HarperBusiness Essentials New York}
}

@article{wishart1928generalised,
  title={The generalised product moment distribution in samples from a normal multivariate population},
  author={Wishart, John},
  journal={Biometrika},
  volume={20},
  number={1/2},
  pages={32--52},
  year={1928},
  publisher={JSTOR}
}

@article{pena2002analisis,
  title={An{\'a}lisis de datos multivariantes. Primera edici{\'o}n, vol 24, Editorial McGraw-Hill},
  author={Pe{\~n}a, D},
  journal={Madrid, Espa{\~n}a},
  year={2002}
}

@article{feral2009largest,
  title={The largest eigenvalues of sample covariance matrices for a spiked population: diagonal case},
  author={F{\'e}ral, Delphine and P{\'e}ch{\'e}, Sandrine},
  journal={Journal of Mathematical Physics},
  volume={50},
  number={7},
  year={2009},
  publisher={AIP Publishing}
}

@article{bloemendal2013limits,
  title={Limits of spiked random matrices I},
  author={Bloemendal, Alex and Vir{\'a}g, B{\'a}lint},
  journal={Probability Theory and Related Fields},
  volume={156},
  number={3},
  pages={795--825},
  year={2013},
  publisher={Springer}
}

@article{sadek2007active,
  title={Active antenna selection in multiuser MIMO communications},
  author={Sadek, Mirette and Tarighat, Alireza and Sayed, Ali H},
  journal={IEEE Transactions on Signal Processing},
  volume={55},
  number={4},
  pages={1498--1510},
  year={2007},
  publisher={IEEE}
}

@article{fyodorov1997statistics,
  title={Statistics of resonance poles, phase shifts and time delays in quantum chaotic scattering: Random matrix approach for systems with broken time-reversal invariance},
  author={Fyodorov, Yan V and Sommers, Hans-J{\"u}rgen},
  journal={Journal of Mathematical Physics},
  volume={38},
  number={4},
  pages={1918--1981},
  year={1997},
  publisher={American Institute of Physics}
}

@article{verbaarschot1994spectrum,
  title={Spectrum of the QCD Dirac operator and chiral random matrix theory},
  author={Verbaarschot, Jacobus},
  journal={Physical Review Letters},
  volume={72},
  number={16},
  pages={2531},
  year={1994},
  publisher={APS}
}

@article{schehr2008exact,
  title={Exact distribution of the maximal height of p vicious walkers},
  author={Schehr, Gr{\'e}gory and Majumdar, Satya N and Comtet, Alain and Randon-Furling, Julien},
  journal={Physical review letters},
  volume={101},
  number={15},
  pages={150601},
  year={2008},
  publisher={APS}
}

@article{baik2005phase,
  title   = {Phase transition of the largest eigenvalue for nonnull complex sample covariance matrices},
  author  = {Baik, Jinho and Ben Arous, G{\'e}rard and P{\'e}ch{\'e}, Sandrine},
  journal = {Annals of Probability},
  volume  = {33},
  number  = {5},
  pages   = {1643--1697},
  year    = {2005},
  publisher = {Institute of Mathematical Statistics}
}

\end{document}